\documentclass[twocolumn]{aastex63}

\newcommand{\rearth}{$R_{\oplus}$}

\received{\today}
\revised{}
\accepted{}
\submitjournal{AJ}

\shorttitle{Confirmation of Two USP Planets around M Dwarfs}
\shortauthors{Hirano et al.}

\begin{document}

\title{Two Bright M Dwarfs Hosting Ultra-Short-Period Super-Earths with Earth-like Compositions\footnote{Based on data collected at Subaru Telescope, which is operated by the National Astronomical Observatory of Japan.}}

\correspondingauthor{Teruyuki Hirano}
\email{teruyuki.hirano@nao.ac.jp}


\author[0000-0003-3618-7535]{Teruyuki Hirano}
\affiliation{Astrobiology Center, 2-21-1 Osawa, Mitaka, Tokyo 181-8588, Japan}
\affiliation{National Astronomical Observatory of Japan, NINS, 2-21-1 Osawa, Mitaka, Tokyo 181-8588, Japan}
\affiliation{Department of Astronomical Science, School of Physical Sciences, The Graduate University for Advanced Studies (SOKENDAI), 2-21-1, Osawa, Mitaka, Tokyo, 181-8588, Japan}

\author[0000-0002-4881-3620]{John H. Livingston}
\affiliation{Department of Astronomy, University of Tokyo, 7-3-1 Hongo, Bunkyo-ku, Tokyo 113-0033, Japan}

\author[0000-0002-4909-5763]{Akihiko Fukui}
\affiliation{Department of Earth and Planetary Science, Graduate School of Science, The University of Tokyo, 7-3-1 Hongo, Bunkyo-ku, Tokyo 113-0033, Japan}
\affiliation{Instituto de Astrof\'{i}sica de Canarias (IAC), 38205 La Laguna, Tenerife, Spain}

\author[0000-0001-8511-2981]{Norio Narita}
\affiliation{Komaba Institute for Science, The University of Tokyo, 3-8-1 Komaba, Meguro, Tokyo 153-8902, Japan}
\affiliation{JST, PRESTO, 3-8-1 Komaba, Meguro, Tokyo 153-8902, Japan}
\affiliation{Astrobiology Center, 2-21-1 Osawa, Mitaka, Tokyo 181-8588, Japan}
\affiliation{Instituto de Astrof\'{i}sica de Canarias (IAC), 38205 La Laguna, Tenerife, Spain}

\author[0000-0002-7972-0216]{Hiroki Harakawa}
\affiliation{Subaru Telescope, 650 N. Aohoku Place, Hilo, HI 96720, USA}

\author[0000-0001-6309-4380]{Hiroyuki Tako Ishikawa}
\affiliation{Astrobiology Center, 2-21-1 Osawa, Mitaka, Tokyo 181-8588, Japan}
\affiliation{National Astronomical Observatory of Japan, 2-21-1 Osawa, Mitaka, Tokyo 181-8588, Japan}

\author{Kohei Miyakawa}
\affiliation{Department of Earth and Planetary Sciences, Tokyo Institute of Technology, Meguro-ku, Tokyo, 152-8551, Japan}

\author{Tadahiro Kimura}
\affiliation{Department of Earth and Planetary Science, Graduate School of Science, The University of Tokyo, 7-3-1 Hongo, Bunkyo-ku, Tokyo 113-0033, Japan}

\author[0000-0002-0998-0434]{Akifumi Nakayama}
\affiliation{Department of Earth and Planetary Science, Graduate School of Science, The University of Tokyo, 7-3-1 Hongo, Bunkyo-ku, Tokyo 113-0033, Japan}

\author[0000-0002-5791-970X]{Naho Fujita}
\affiliation{Department of Astronomy, Kyoto University, Kitashirakawa-Oiwake-cho, Sakyo-ku, Kyoto 606-8502, Japan}

\author[0000-0003-4676-0251]{Yasunori Hori}
\affiliation{Astrobiology Center, 2-21-1 Osawa, Mitaka, Tokyo 181-8588, Japan}
\affiliation{National Astronomical Observatory of Japan, NINS, 2-21-1 Osawa, Mitaka, Tokyo 181-8588, Japan}

\author[0000-0002-3481-9052]{Keivan G.\ Stassun}
\affiliation{Department of Physics and Astronomy, Vanderbilt University, 6301 Stevenson Center Ln., Nashville, TN 37235, USA}
\affiliation{Department of Physics, Fisk University, 1000 17th Avenue North, Nashville, TN 37208, USA}

\author[0000-0001-6637-5401]{Allyson Bieryla} 
\affiliation{Center for Astrophysics \textbar \ Harvard \& Smithsonian, 60 Garden Street, Cambridge, MA 02138, USA}

\author[0000-0001-9291-5555]{Charles Cadieux}
\affiliation{\href{http://www.exoplanetes.ca/}{Institute for Research on Exoplanets} (IREx), Universit\'e de Montr\'eal, D\'epartement de Physique, C.P.~6128 Succ. Centre-ville, Montr\'eal, QC H3C~3J7, Canada}

\author[0000-0002-5741-3047]{David R. Ciardi} 
\affiliation{NASA Exoplanet Science Institute, Caltech/IPAC, Mail Code 100-22, 1200 E. California Blvd., Pasadena, CA 91125, USA}

\author[0000-0001-6588-9574]{Karen A.\ Collins}
\affiliation{Center for Astrophysics \textbar \ Harvard \& Smithsonian, 60 Garden Street, Cambridge, MA 02138, USA}

\author[0000-0002-5658-5971]{Masahiro Ikoma}
\affiliation{Department of Earth and Planetary Science, Graduate School of Science, The University of Tokyo, 7-3-1 Hongo, Bunkyo-ku, Tokyo 113-0033, Japan}

\author[0000-0001-7246-5438]{Andrew Vanderburg}
\affiliation{Department of Astronomy, The University of Wisconsin-Madison, Madison, WI 53706, USA}

\author[0000-0001-7139-2724]{Thomas Barclay}
\affiliation{NASA Goddard Space Flight Center, 8800 Greenbelt Road, Greenbelt, MD 20771, USA}
\affiliation{University of Maryland, Baltimore County, 1000 Hilltop Circle, Baltimore, MD 21250, USA}

\author[0000-0002-9314-960X]{C. E. Brasseur}
\affiliation{Space Telescope Science Institute, 3700 San Martin Drive, Baltimore, MD, 21218, USA}

\author[0000-0002-6424-3410]{Jerome P. de Leon}
\affiliation{Department of Astronomy, Graduate School of Science, The University of Tokyo, 7-3-1 Hongo, Bunkyo-ku, Tokyo 113-0033, Japan}

\author{John~P.~Doty}
\affiliation{Noqsi Aerospace Ltd., 15 Blanchard Avenue, Billerica, MA 01821, USA}

\author[0000-0001-5485-4675]{Ren\'e Doyon}
\affil{\href{http://www.exoplanetes.ca/}{Institute for Research on Exoplanets} (IREx), Universit\'e de Montr\'eal, D\'epartement de Physique, C.P.~6128 Succ. Centre-ville, Montr\'eal, QC H3C~3J7, Canada}

\author[0000-0002-2341-3233]{Emma Esparza-Borges}
\affiliation{Instituto de Astrof\'\i sica de Canarias (IAC), 38205 La Laguna, Tenerife, Spain}
\affiliation{Departamento de Astrof\'\i sica, Universidad de La Laguna (ULL), 38206, La Laguna, Tenerife, Spain}

\author[0000-0002-9789-5474]{Gilbert A. Esquerdo} 
\affiliation{Center for Astrophysics \textbar \ Harvard \& Smithsonian, 60 Garden Street, Cambridge, MA 02138, USA}

\author[0000-0001-9800-6248]{Elise Furlan}
\affiliation{NASA Exoplanet Science Institute, Caltech/IPAC, Mail Code 100-22, 1200 E. California Blvd., Pasadena, CA 91125, USA}

\author[0000-0002-5258-6846]{Eric Gaidos}
\affiliation{Department of Earth Sciences, University of Hawai'i at M\"{a}noa, Honolulu, HI 96822, USA}

\author{Erica J. Gonzales}
\affiliation{Department of Physics, University of Notre Dame, 225 Nieuw-
land Science Hall, Notre Dame IN 46556, USA}
\affiliation{University of California, Santa Cruz, 1156 High Street, Santa
Cruz CA 95065, USA}

\author[0000-0003-0786-2140]{Klaus Hodapp}
\affiliation{University of Hawaii, Institute for Astronomy, 640 N. Aohoku Place, Hilo, HI 96720, USA}

\author[0000-0002-2532-2853]{Steve B. Howell}
\affiliation{NASA Ames Research Center, Moffett Field, CA 94035, USA}


\author[0000-0002-6480-3799]{Keisuke Isogai}
\affiliation{Okayama Observatory, Kyoto University, 3037-5 Honjo, Kamogatacho, Asakuchi, Okayama 719-0232, Japan}
\affiliation{Department of Multi-Disciplinary Sciences, Graduate School of Arts and Sciences, The University of Tokyo, 3-8-1 Komaba, Meguro, Tokyo 153-8902, Japan}

\author{Shane Jacobson}
\affiliation{University of Hawaii, Institute for Astronomy, 640 N. Aohoku Place, Hilo, HI 96720, USA}

\author[0000-0002-4715-9460]{Jon M. Jenkins}
\affiliation{NASA Ames Research Center, Moffett Field, CA 94035, USA}

\author[0000-0002-4625-7333]{Eric L.\ N.\ Jensen}
\affiliation{Department of Physics \& Astronomy, Swarthmore College, Swarthmore PA 19081, USA}

\author[0000-0003-1205-5108]{Kiyoe Kawauchi}
\affiliation{Department of Earth and Planetary Science, Graduate School of Science, The University of Tokyo, 7-3-1 Hongo, Bunkyo-ku, Tokyo 113-0033, Japan}


\author[0000-0001-6181-3142]{Takayuki Kotani}
\affiliation{Astrobiology Center, 2-21-1 Osawa, Mitaka, Tokyo 181-8588, Japan}
\affiliation{National Astronomical Observatory of Japan, NINS, 2-21-1 Osawa, Mitaka, Tokyo 181-8588, Japan}
\affiliation{Department of Astronomical Science, School of Physical Sciences, The Graduate University for Advanced Studies (SOKENDAI), 2-21-1, Osawa, Mitaka, Tokyo, 181-8588, Japan}

\author[0000-0002-9294-1793]{Tomoyuki Kudo}
\affiliation{Subaru Telescope, 650 N. Aohoku Place, Hilo, HI 96720, USA}

\author{Seiya Kurita}
\affiliation{Department of Earth and Planetary Science, Graduate School of Science, The University of Tokyo, 7-3-1 Hongo, Bunkyo-ku, Tokyo 113-0033, Japan}

\author{Takashi Kurokawa}
\affiliation{Astrobiology Center, 2-21-1 Osawa, Mitaka, Tokyo 181-8588, Japan}
\affiliation{Institute of Engineering, Tokyo University of Agriculture and Technology, 2-24-16, Nakacho, Koganei, Tokyo, 184-8588, Japan}

\author[0000-0001-9194-1268]{Nobuhiko Kusakabe}
\affiliation{Astrobiology Center, 2-21-1 Osawa, Mitaka, Tokyo 181-8588, Japan}
\affiliation{National Astronomical Observatory of Japan, 2-21-1 Osawa, Mitaka, Tokyo 181-8588, Japan}

\author[0000-0002-4677-9182]{Masayuki Kuzuhara}
\affiliation{Astrobiology Center, 2-21-1 Osawa, Mitaka, Tokyo 181-8588, Japan}
\affiliation{National Astronomical Observatory of Japan, NINS, 2-21-1 Osawa, Mitaka, Tokyo 181-8588, Japan}

\author[0000-0002-6780-4252]{David Lafreni\` ere}
\affil{\href{http://www.exoplanetes.ca/}{Institute for Research on Exoplanets} (IREx), Universit\'e de Montr\'eal, D\'epartement de Physique, C.P.~6128 Succ. Centre-ville, Montr\'eal, QC H3C~3J7, Canada}

\author[0000-0001-9911-7388]{David W. Latham} 
\affiliation{Center for Astrophysics \textbar \ Harvard \& Smithsonian, 60 Garden Street, Cambridge, MA 02138, USA}

\author[0000-0001-8879-7138]{Bob Massey}
\affil{Villa '39 Observatory, Landers, CA 92285, USA}

\author[0000-0003-1368-6593]{Mayuko Mori}
\affiliation{Department of Astronomy, Graduate School of Science, The University of Tokyo, 7-3-1 Hongo, Bunkyo-ku, Tokyo 113-0033, Japan}

\author{Felipe Murgas}
\affiliation{Instituto de Astrof\'isica de Canarias (IAC), E-38205 La Laguna, Tenerife, Spain}
\affiliation{Departamento de Astrof\'isica, Universidad de La Laguna (ULL), E-38206 La Laguna, Tenerife, Spain}

\author[0000-0001-9326-8134]{Jun Nishikawa}
\affiliation{National Astronomical Observatory of Japan, NINS, 2-21-1 Osawa, Mitaka, Tokyo 181-8588, Japan}
\affiliation{Department of Astronomical Science, School of Physical Sciences, The Graduate University for Advanced Studies (SOKENDAI), 2-21-1, Osawa, Mitaka, Tokyo, 181-8588, Japan}
\affiliation{Astrobiology Center, 2-21-1 Osawa, Mitaka, Tokyo 181-8588, Japan}

\author[0000-0003-1510-8981]{Taku Nishiumi}
\affiliation{Department of Astronomical Science, School of Physical Sciences, The Graduate University for Advanced Studies (SOKENDAI), 2-21-1, Osawa, Mitaka, Tokyo, 181-8588, Japan}
\affiliation{Astrobiology Center, 2-21-1 Osawa, Mitaka, Tokyo 181-8588, Japan}

\author[0000-0002-5051-6027]{Masashi Omiya}
\affiliation{Astrobiology Center, 2-21-1 Osawa, Mitaka, Tokyo 181-8588, Japan}
\affiliation{National Astronomical Observatory of Japan, NINS, 2-21-1 Osawa, Mitaka, Tokyo 181-8588, Japan}

\author[0000-0001-8120-7457]{Martin Paegert}
\affiliation{Center for Astrophysics \textbar \ Harvard \& Smithsonian, 60 Garden Street, Cambridge, MA 02138, USA}

\author{Enric Palle}
\affiliation{Instituto de Astrof\'\i sica de Canarias (IAC), 38205 La Laguna, Tenerife, Spain}
\affiliation{Departamento de Astrof\'\i sica, Universidad de La Laguna (ULL), 38206, La Laguna, Tenerife, Spain}

\author[0000-0001-5519-1391]{Hannu Parviainen}
\affiliation{Instituto de Astrof\'\i sica de Canarias (IAC), 38205 La Laguna, Tenerife, Spain}
\affiliation{Departamento de Astrof\'\i sica, Universidad de La Laguna (ULL), 38206, La Laguna, Tenerife, Spain}

\author[0000-0002-8964-8377]{Samuel N. Quinn} 
\affiliation{Center for Astrophysics \textbar \ Harvard \& Smithsonian, 60 Garden Street, Cambridge, MA 02138, USA}

\author{George\,R.~Ricker} 
\affiliation{Department of Physics and Kavli Institute for Astrophysics and Space Research, Massachusetts Institute of Technology, 77 Massachusetts Avenue, Cambridge, MA 02139, USA}

\author[0000-0001-8227-1020]{Richard P. Schwarz}
\affiliation{Patashnick Voorheesville Observatory, Voorheesville, NY 12186, USA}

\author[0000-0002-6892-6948]{Sara Seager}
\affiliation{Department of Physics and Kavli Institute for Astrophysics and Space Research, Massachusetts Institute of Technology, 77 Massachusetts Avenue, Cambridge, MA 02139, USA}
\affiliation{Department of Earth, Atmospheric and Planetary Sciences, Massachusetts Institute of Technology, Cambridge, MA 02139, USA}
\affiliation{Department of Aeronautics and Astronautics, Massachusetts Institute of Technology, Cambridge, MA 02139, USA}


\author[0000-0002-6510-0681]{Motohide Tamura}
\affiliation{Department of Astronomy, Graduate School of Science, The University of Tokyo, 7-3-1 Hongo, Bunkyo-ku, Tokyo 113-0033, Japan}
\affiliation{Astrobiology Center, 2-21-1 Osawa, Mitaka, Tokyo 181-8588, Japan}
\affiliation{National Astronomical Observatory of Japan, NINS, 2-21-1 Osawa, Mitaka, Tokyo 181-8588, Japan}

\author[0000-0002-1949-4720]{Peter Tenenbaum}
\affiliation{SETI Institute / NASA Ames Research Center, Moffett Field, CA 94035, USA}

\author[0000-0003-2887-6381]{Yuka Terada}
\affiliation{Institute of Astronomy and Astrophysics, Academia Sinica, P.O. Box 23-141, Taipei 10617, Taiwan, R.O.C.}
\affiliation{Department of Astrophysics, National Taiwan University, Taipei 10617, Taiwan, R.O.C.}


\author[0000-0001-6763-6562]{Roland K. Vanderspek}
\affiliation{Department of Physics and Kavli Institute for Astrophysics and Space Research, Massachusetts Institute of Technology, 77 Massachusetts Avenue, Cambridge, MA 02139, USA}

\author[0000-0003-4018-2569]{S\'ebastien Vievard}
\affiliation{Astrobiology Center, 2-21-1 Osawa, Mitaka, Tokyo 181-8588, Japan}
\affiliation{Subaru Telescope, 650 N. Aohoku Place, Hilo, HI 96720, USA}

\author[0000-0002-7522-8195]{Noriharu Watanabe}
\affiliation{Department of Astronomical Science, School of Physical Sciences, The Graduate University for Advanced Studies (SOKENDAI), 2-21-1, Osawa, Mitaka, Tokyo, 181-8588, Japan}
\affiliation{Astrobiology Center, 2-21-1 Osawa, Mitaka, Tokyo 181-8588, Japan}

\author[0000-0002-4265-047X]{Joshua N.\ Winn}
\affiliation{Department of Astrophysical Sciences, Princeton University, 4 Ivy Lane, Princeton, NJ 08544, USA}





\begin{abstract}
We present observations of two bright M dwarfs (TOI-1634 and TOI-1685: $J=9.5-9.6$) hosting ultra-short period (USP) planets, 
identified by the TESS mission. 
The two stars are similar in temperature, mass, and radius  ($T_\mathrm{eff}\,\approx\,3500$ K, $M_\star\,\approx\,0.45-0.46\,M_\odot$, and $R_\star\approx 0.45-0.46\,R_\odot$), and the planets are both super-Earth-sized ($1.25\,R_\oplus<R_p<2.0\,R_\oplus$). 
For both systems, 
light curves from the ground-based photometry exhibit planetary transits, whose depths
are consistent with those by the TESS photometry. 
We also refine the transit ephemerides based on the ground-based photometry, finding the orbital periods of $P=0.9893436\pm0.0000020$ day and $P=0.6691416\pm0.0000019$ day for TOI-1634b and TOI-1685b, respectively. 
Through intensive radial velocity (RV) observations using IRD on the Subaru 8.2m telescope, we confirm the planetary nature of the TOIs, and measure their masses: 
$10.14\pm0.95\,M_\oplus$ and $3.43\pm0.93\,M_\oplus$
for TOI-1634b and TOI-1685b, respectively, when the observed RVs are fitted with a single-planet circular-orbit model. Combining those with the planet radii of 
$R_p=1.749\pm 0.079\,R_\oplus$ (TOI-1634b) and $1.459\pm0.065\,R_\oplus$ (TOI-1685b), 
we find that both USP planets have mean densities consistent with an Earth-like internal composition, which is typical for small USP planets. TOI-1634b is 
currently the 
most massive USP planet in this category, and it resides near the radius valley, which makes it a benchmark planet in the context of 
discussing the size limit of rocky planet cores as well as testing 
the formation scenarios for USP planets. 
Excess scatter in the RV residuals for TOI-1685 suggests the presence of a possible secondary planet or unknown activity/instrumental noise in the RV data, but further observations are required to check those possibilities. 
\end{abstract}



\section{Introduction} \label{sec:intro}

Ultra-short-period (USP) planets refer to a class of exoplanets (usually with radii smaller than $2\,R_\oplus$) with periods less than 1.0 day. 
Since the earliest examples were discovered back in the late 2000's \citep{2006Natur.443..534S, 2009A&A...506..287L}, 
more than 100 such USP planets have been reported to date.
Recent statistical studies have shown that USP planets are as rare as hot Jupiters, and their occurrence rate seems to depend on the host star's type;
the occurrence rate is estimated as $1.1\pm0.4\,\%$ for M dwarfs, but it falls to $0.15\pm 0.05\,\%$
for F dwarfs \citep{2018NewAR..83...37W}. 
USP planets are often found in multi-planet systems, but the period ratios and mutual inclinations for the adjacent planet pairs are reported to be different from those for longer-period planets ($P>1$ day) in multi-planet systems \citep{2013ApJ...774L..12S, 2018NewAR..83...37W}. 
It had been proposed that USP planets are remnant rocky/iron cores of hot Jupiters that 
have experienced dissipations of their gaseous envelopes due to photoevaporation or 
Roche lobe overflow \citep[e.g.,][]{2010A&A...516A..20V,2013ApJ...779..165J, 2016CeMDA.126..227J, 2017ApJ...846L..13K}, 
but this hypothesis turned out unlikely 
after \citet{2018NewAR..83...37W} found that
stars hosting USP planets have a different metallicity distribution from that of the 
hot-Jupiter hosting stars; 
while hot Jupiter are preferentially hosted by metal-rich stars with their
occurrence rate rising with the third or fourth power of metallicity
\citep{2018AJ....155...89P}, 
the metallicities of USP-planet hosts have a broader distribution with 
its peak around $\mathrm{[Fe/H]=0.0}$ \citep{2017AJ....154...60W}, 
which is more similar to Kepler multi-planet systems (without hot Jupiters).

The origin of USP planets have been discussed in the literature, and 
almost all scenarios require some inward planet migration as opposed to in-situ
formation, since the observed locations of USP planets are well inside 
the dust sublimation radius of the protoplanetary disk. 
USP planets typically have circularized orbits. 
Tidal interactions between the star and the close-in planet are likely responsible for the low eccentricities of USP planets. While tides may have also played an important role in the formation of USP planets, tidal dissipation alone is unable to 
generate USP planets with a reasonable assumption for the tidal quality factor 
\citep[e.g.,][]{2010ApJ...723..285H, 2019AJ....157..180P}. 
To explain the presence of USP planets, ``high-eccentricity migration" scenarios
among close-in planets were proposed \citep[e.g.,][]{2010ApJ...724L..53S}, 
which are miniature versions of the possible formation channel for hot Jupiters. 
Recently, alternative scenarios have been suggested to explain the observed
eccentricity and mutual inclination of USP planets. 
\citet{2019MNRAS.488.3568P} 
investigated the low-eccentricity tidal migration induced by secular planet-planet interactions, 
finding that their scenario can produce the USP population largely consistent with the observed Kepler multi-planet systems. 
More recently, \citet{2020ApJ...905...71M} proposed a new channel to form USP planets 
through a non-zero planetary obliquity driving tidal dissipations. 
Their scenario also predicts the properties of USP planets that are broadly consistent with the observed features such as the period ratios and occurrence rate trends with stellar type.

In order to corroborate or refute those hypotheses for the origin of USP 
planets, we should compare the prediction of individual 
theoretical models with the observed properties of the systems including USP planets, 
such as the dependence on the stellar type and the period and mass ratios of the neighboring planets in multi-planet systems. 
However, the number of ``well characterized" USP planets with precisely measured masses and radii is still limited 
to date. In particular, only two USP planets around
M dwarfs (LTT 3780 and GJ 1252) have precise mass measurements 
\citep{2020AJ....160....3C, 2020A&A...642A.173N, 2020ApJ...890L...7S}. 
Radial velocity (RV) follow-up observations are important for USP planets not only 
in terms of confirmation of the candidates, but also for constraining the bulk compositions of the planets, which shed some light on the origin and evolution of
USP planets. Moreover, RV monitorings allow for
the search for additional planets responsible for the formation of inner USP planets, which may not be transiting in the 
presence of significant mutual inclinations between the planets \citep[e.g., $\gtrsim 5^\circ$ in][]{2018ApJ...864L..38D}. 

In this paper, we report on the validation and confirmation of new USP planets around 
two M dwarfs, whose transits were identified by the TESS mission
\citep{2015JATIS...1a4003R}. 
Since TESS started its scientific operation in 2018, the spacecraft participated in the 
search for USP planets. As of 2021 February, 151 USP planet candidates were
reported as TESS Objects of Interest 
\citep[TOI's;][]{2021arXiv210312538G} by the mission 
(excluding the ones flagged as ``False Positive (FP)"), and 31 of them are orbiting 
M dwarfs (the effective temperature $T_\mathrm{eff}<4000$ K). 
Our targets are TOI-1634 and TOI-1685, which are similar in the stellar $T_\mathrm{eff}$, mass $M_\star$, and radius $R_\star$, hosting super-Earth-sized USP planet candidates
according to the TESS Input Catalog \citep[TIC;][]{2019AJ....158..138S}. 
As the properties are shown in Table \ref{hyo1}, those two targets are both relatively 
bright M dwarfs as hosts of transiting-planet candidates (i.e., both are close to Earth), and thus would become
excellent targets for future characterizations once validated. 
With the goal of confirming those candidates as well as deriving precise and accurate 
system parameters, we conducted follow-up observations for those systems including ground-based transit photometry and precise RV observations.

The rest of the paper is organized as follows. 
Section \ref{sec:obs} presents the details of TESS transit photometry as well as our 
imaging/photometric and spectroscopic follow-up observations. 
We describe the analyses of the new data and their results in Section \ref{sec:ana}, 
providing our new estimations of the system parameters. 
In Section \ref{sec:discussion}, we will discuss the physical properties of new planets
as well as the possibility of future follow-up studies. 
Finally, our brief summary is given in Section \ref{sec:summary}.

\begin{table}[t]
\centering
\caption{Stellar Parameters of TOI-1634 and TOI-1685}\label{hyo1}
\begin{tabular}{lcc}
\hline\hline
Parameter & TOI-1634 & TOI-1685 \\\hline
\multicolumn{2}{l}{\bf (Literature Values)} & \\
TIC & 201186294 & 28900646\\
{\bf 2MASS ID} & J03453363+3706438 & J04342248+4302148\\
$\alpha$ (J2000)$^a$ & 03:45:33.641 & 04:34:22.495 \\
$\delta$ (J2000)$^a$ & +37:06:43.999 & +43:02:14.692 \\
$\mu_\alpha\cos\delta$ (mas yr$^{-1}$)$^a$ & $81.348\pm 0.020$ & $37.762\pm 0.022$ \\
$\mu_\delta$ (mas yr$^{-1}$)$^a$ & $13.548\pm 0.015$ & $-87.062\pm 0.018$ \\
parallax (mas)$^a$ & $28.5123 \pm 0.0184$ &  $26.5893 \pm 0.0192$ \\
Gaia $G$ (mag)$^a$ & $12.1965 \pm 0.0003$ & $12.2956 \pm 0.0003$ \\
TESS $T$ (mag)$^b$ & $11.0136 \pm 0.0073$ & $11.1117 \pm 0.0073$ \\
$J$ (mag)$^c$ & $9.484 \pm 0.021$ & $9.616 \pm 0.022$ \\
$H$ (mag)$^c$ & $8.847 \pm 0.021$ & $9.005 \pm 0.023$ \\
$K$ (mag)$^c$ & $8.600 \pm 0.014$ & $8.758 \pm 0.020$ \\\hline
\multicolumn{2}{l}{\bf (Derived Values)} & \\
$d$ (pc) & $35.072 \pm 0.023$ & $37.609 \pm 0.027$\\
$T_\mathrm{eff}$ (K) & $3472 \pm 70$ & $3461 \pm 70$ \\
$U$ (km s$^{-1}$) & $9.58 \pm 0.45$ & $35.53 \pm 0.47$ \\
$V$ (km s$^{-1}$) & $-13.81 \pm 0.19$ & $-29.82 \pm 0.17$ \\
$W$ (km s$^{-1}$) & $14.08 \pm 0.12$ & $-3.14 \pm 0.03$ \\
$[\mathrm{Fe/H}]$ (dex) & $0.19\pm0.12$ & $0.14\pm 0.12$ \\
$[\mathrm{Na/H}]$ (dex) & $0.20\pm0.14$ & $0.24\pm 0.14$ \\
$[\mathrm{Mg/H}]$ (dex) & $0.38\pm0.18$ & $0.45\pm 0.19$ \\
$[\mathrm{Si/H}]$ (dex) & $0.77\pm0.31$ & $0.55\pm 0.30$ \\
$[\mathrm{Ca/H}]$ (dex) & $0.19\pm0.12$ & $0.21\pm 0.13$ \\
$[\mathrm{Ti/H}]$ (dex) & $0.58\pm0.21$ & $0.71\pm 0.24$ \\
$[\mathrm{Cr/H}]$ (dex) & $0.29\pm0.12$ & $0.29\pm 0.12$\\
$[\mathrm{Mn/H}]$ (dex) & $0.32\pm0.17$ & $0.35\pm 0.17$\\
$\log g$ (cgs) & $4.787\pm 0.027$ & $4.778 \pm 0.026$  \\
$M_\star$ ($M_\odot$) & $0.451\pm 0.015$ & $0.460\pm 0.011$ \\
$R_\star$ ($R_\odot$) & $0.450\pm 0.016$ & $0.459 \pm 0.013$ \\
$\rho_\star$ (g cm$^{-3}$) & $6.98_{-0.63}^{+0.70}$ & $6.70_{-0.55}^{+0.61}$ \\
$F_{\rm bol}$ (erg~s$^{-1}$~cm$^{-2}$) & $(7.64\pm 0.27)\times 10^{-10}$  & $(6.65\pm 0.15)\times 10^{-10}$ \\
$L_\star$ ($L_\odot$) & $	0.0264_{-0.0027}^{+0.0030}$ & $0.0271_{-0.0026}^{+0.0028}$ \\
\hline
\end{tabular}
\\References: 
$a$) \citet{2020arXiv201201533G}, 
$b$) \citet{2019AJ....158..138S}, 
$c$) \citet{2006AJ....131.1163S}
\end{table}

\section{Observations and Data Reduction} \label{sec:obs}

\subsection{Photometry} \label{sec:obs_im}

\subsubsection{TESS Photometry}
\begin{figure}
    \centering
    \includegraphics[width=0.4\textwidth]{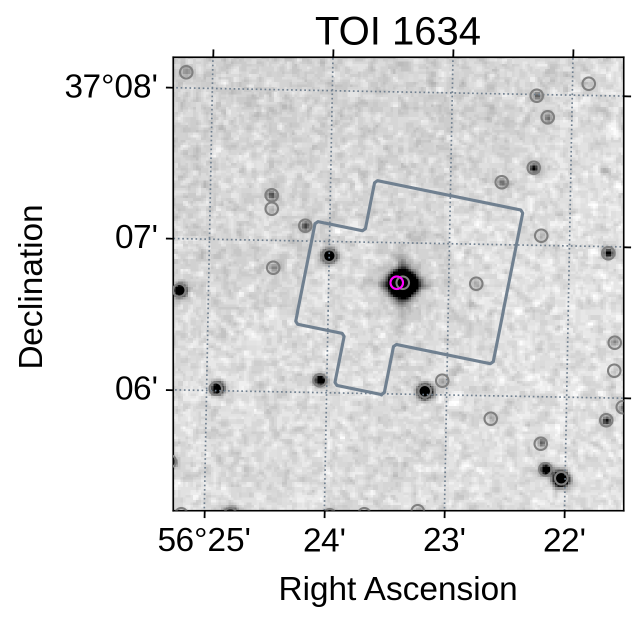}
    \includegraphics[width=0.4\textwidth]{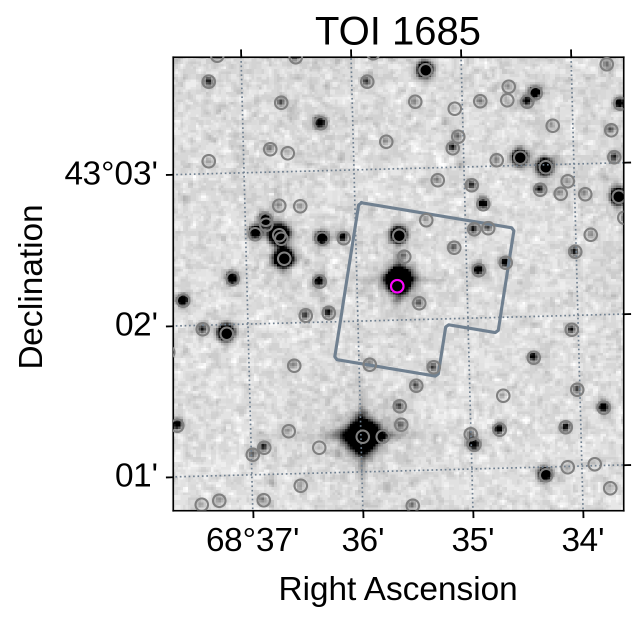}
    \caption{TESS photometric apertures and $3\arcmin\times3\arcmin$ archival images for TOI-1634 (top) and TOI-1685 (bottom). The archival images are scanned photographic plates using the RG610 filter and the IIIaF emulsion, which were originally obtained as part of the POSSII-F survey on September 18, 1988 (TOI-1634) and October 6, 1989 (TOI-1685). The Gaia DR2 positions (epoch J2015.5) of the target stars are indicated by magenta circles, and other sources by gray circles.}
    \label{fig:apertures}
\end{figure}

TESS observed TOI-1634 and TOI-1685 at a 2 min cadence in Sectors 18 and 19, respectively. The observations were conducted from UT 2019 November 03 to 2019 December 23, resulting in photometry spanning approximately 27 days for each target, with gaps of about one day for data downlink in the middle of each observing sequence. Near the beginning of Sector 18 there is an additional 6.2 hour data gap due to the instrument being shut down for Earth eclipse. Light curves were produced by the Science Processing Operations Center (SPOC) photometry pipeline \citep{Jenkins2016} using the apertures shown in Figure~\ref{fig:apertures}. For our transit analyses, we used the {\tt PDCSAP} light curves produced by the SPOC pipeline \citep{2012PASP..124..985S,2012PASP..124.1000S,2014PASP..126..100S}. An error in the SPOC pipeline resulted in over-subtraction of the sky background, causing fractional changes (e.g. transits) in the light curves of TOI-1634 and TOI-1685 to be artificially deeper by 2.2\% and 2.9\%, respectively (Jon Jenkins, private communication). To account for this, we applied a correction to the $R_p/R_\star$ values from our fits to the TESS data before combining them with our ground-based photometric measurements (see Section~\ref{sec:photometry}); we note the effect is smaller than the uncertainty of the $R_p/R_\star$ values derived from the TESS light curves and has negligible impact on the final values. The SPOC pipeline applies a photometric dilution correction based on the {\tt CROWDSAP} metric, which we independently confirmed by computing dilution values based on Gaia DR2 magnitudes (approximating $G_{RP}$ as the TESS bandpass, and assuming a full width at half maximum (FWHM) of 25\arcsec). For TOI-1634 there are two significantly contaminating sources in the aperture (Gaia DR2 IDs 223158499176634112 and 223158808416782208), which are 2.9 and 4.8 magnitudes fainter in the $G_{RP}$ band, respectively. For TOI-1685 there are three significantly contaminating sources (Gaia DR2 IDs 252366613254979328, 252366578895244672, and 252366578895245696), which are 3.7, 6.0, and 6.5 mag fainter in $G_{RP}$, respectively; an additional source (Gaia DR2 ID 252363589598010240) located just outside and to the south of the aperture is 0.25 magnitudes brighter than TOI-1685 and thus also significantly contaminating despite contributing less than 10\% of its flux.

TOI-1634 has a resolved companion star separated by $2\farcs5$ from the primary star (Gaia DR2 ID 223158499176634112), 
with a TESS magnitude of $14.368\pm 0.010$ mag (i.e., about $3.3$ mag fainter
than TOI-1634). The Gaia astrometry indicates the companion star has the parallax of $28.62\pm 0.11$ mas 
and proper motions of $\mu_\alpha\cos\delta=80.64\pm 0.13$ mas yr$^{-1}$ and 
$\mu_\delta =14.539\pm 0.091$ mas yr$^{-1}$, respectively \citep{2020arXiv201201533G}, 
suggesting that TOI-1634 and the companion star share almost the same parallax and 
common proper motions. Thus, they are likely bound to each other, which was also 
reported in the visual-binary catalog for TOI's \citep{2020AN....341..996M} as well as the more recent catalog by \citet{2021MNRAS.tmp..394E} based on Gaia EDR3. 
Light curve dilutions due to this companion star are taken into account when we
perform the light curve analyses. The impact of the companion on the estimation of
the stellar properties as well as the long-term RV drift for TOI-1634 will be
discussed in Sections \ref{sec:parameters} and \ref{sec:RV}. 
Other than this companion star, no stars were identified within $1^\prime$ in the Gaia EDR3 catalog having proper motions in common with TOI-1634 and TOI-1685. 

The signature of TOI-1634.01 was initially detected by the TESS SPOC in a transiting planet search of sector 18 that occurred UT on 2019 December 12, yielding a 1.8\rearth\ planet in a 0.98933 day orbit about its host star. The signal was detected at 10.6$\sigma$ with an adaptive, noise-compensating matched filter 
\citep{2002ApJ...575..493J, 2010SPIE.7740E..0DJ, 2020ksci.rept....9J},
passed all the diagnostic tests performed and published in the resulting Data Validation reports and was fitted with a limb-darkened transit model 
\citep{Twicken:DVdiagnostics2018, Li:DVmodelFit2019}
. These included tests for eclipsing binaries, such as an odd/even depth test, a weak secondary test, and a ghost diagnostic test. The difference imaging centroid test showed that the source of the transit signature was consistent with the target star, TIC 201186294, with a measured offset from the target star of $8.1 \pm 2.9\arcsec$ (we take 3 sigma as the confusion radius). The SPOC pipeline search removed the signature of TOI-1634.01 from the light curve and performed a search for additional transit signatures, which were not found. An alert for TOI-1634.01 was issued by the TESS Science Office (TSO) on UT 2020 January 14. 

The signature of TOI-1685.01 was detected by the SPOC pipeline in a transiting planet search of Sector 19 that occurred on UT 2020 January 17, resulting in a 1.47\rearth\ planet in a 0.6669 day orbit. This transit signature passed all the diagnostic tests performed and reported in the Data Validation reports archived to MAST and the TSO alerted the community to this planet candidate on UT 2020 January 30. The difference imaging centroid test showed that the source of the transit signature was consistent with the target star, TIC 28900646, with a measured offset from the target star of $2.79 \pm 2.66\arcsec$. As was done for TOI-1634, the SPOC pipeline removed the signature of TOI-1685.01 from the light curve and performed a search for additional transit signatures, which were not found. We note that these difference imaging centroid measurements are complementary to the high resolution imaging reported in Section~2.2, which is limited to separations of 1.2\arcsec\ and 3.0\arcsec\ from each target.

\begin{figure*}
    \centering
    \includegraphics[width=0.9\textwidth]{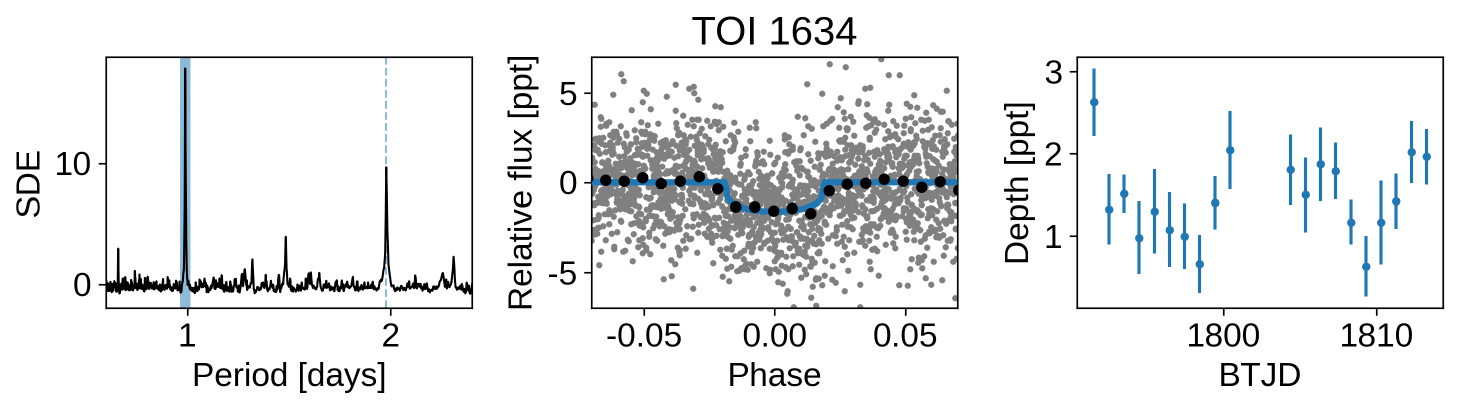}
    \includegraphics[width=0.9\textwidth]{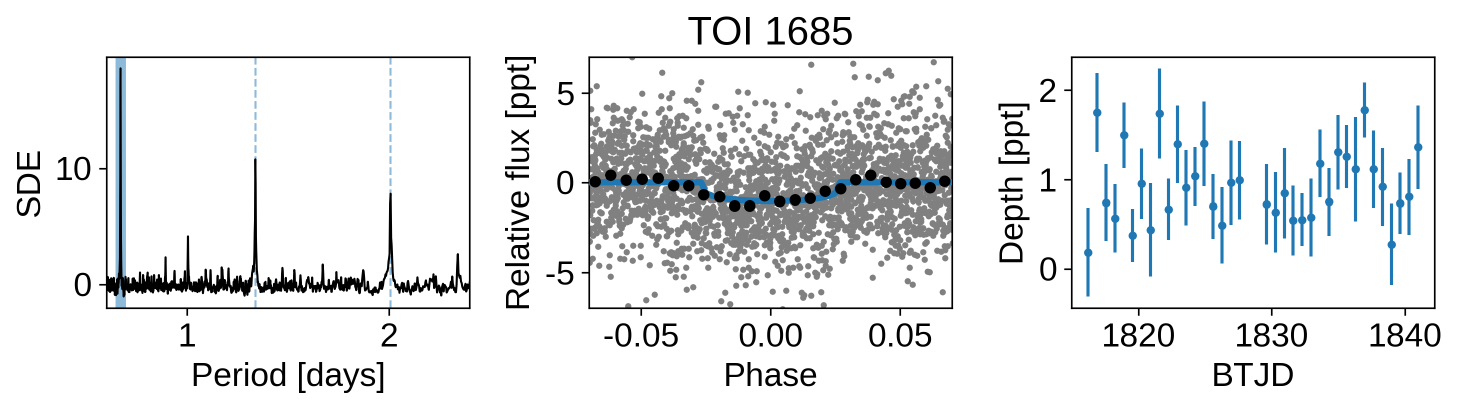}
    \caption{TLS transit signal detections for TOI-1634 (top) and TOI-1685 (bottom). The left panels show SDE vs orbital period; the middle panels show the data folded on the detected period with the TLS model in blue, binned data in black; the right panels show the individual transit depths.}
    \label{fig:tls}
\end{figure*}

We independently confirmed the transit signals of each planet candidate using a 2$\mathrm{nd}$ order polynomial Savitzy-Golay filter to remove stellar variability and instrumental systematics from each light curve, then used the transit least-squares algorithm \citep[TLS;][]{2019A&A...623A..39H}\footnote{\url{https://transitleastsquares.readthedocs.io/en/latest/index.html}} to search them for transit signals, resulting in a signal detection efficiency (SDE) of 17.9, orbital period of $0.989 \pm 0.003$ days, and transit depth of 1.6 parts per thousand (ppt) for TOI-1634.01, and SDE of 18.6, orbital period of $0.669 \pm 0.001$ days, and transit depth of 1.0 ppt for TOI-1685.01. We subtracted each signal and repeated the transit search, but no additional transit signals with SDE above 10 were found in either light curve. TLS also reports the approximate depths of each individual transit; we note that these transit depths and uncertainties are useful for diagnostic purposes only, as they are simplistically determined from the mean and standard deviation of the in-transit flux. The depths of the odd transits are within 0.5$\sigma$ of the even transits for both signals, suggesting a low probability of either signal being caused by an eclipsing binary at twice the detected period. These signals are consistent with those reported by the TESS team on ExoFOP-TESS\footnote{\url{https://exofop.ipac.caltech.edu/tess/}}. The TLS detections are shown in Figure~\ref{fig:tls}.

\subsubsection{Okayama 188~cm / MuSCAT Photometry}

We observed four transits of TOI-1685.01 on UT 2020 November 24, UT 2021 January 10, UT 2021 January 12, and UT 2021 January 14, using the multiband imager MuSCAT \citep{2015JATIS...1d5001N} mounted on the 188~cm telescope at Okayama Astro-Complex in Japan. MuSCAT has three channels for $g$-, $r$-, and $z_s$ bands, enabling three-band simultaneous imaging observations. Each channel is equipped with a 1024 $\times$ 1024 pixel CCD camera with a pixel scale of 0\farcs36 pixel$^{-1}$, which provides a field of view (FOV) of 6\farcm1 square. We observed the target field with exposure times of 6 to 30 sec depending on the band and sky condition. The obtained images were corrected for dark and flat in a standard manner, and aperture photometry was performed by a custom-built photometry pipeline \citep{2011PASJ...63..287F} to produce normalized light curves, in which the combinations of comparison stars and aperture radius were optimized such that the light curve dispersion was minimized. The adopted aperture radius ranges from 8 to 14 pixels (from 2\farcs9 to 5\farcs1) depending on the band and night.

\subsubsection{IAC 1.52m / MuSCAT2 Photometry}

We observed five transits of TOI-1634.01 on UT 2020 February 7, UT 2020 February 10, UT 2020 February 11, UT 2021 February 14, and UT 2021 February 16 using the multiband imager MuSCAT2 \citep{2019JATIS...5a5001N} mounted on the 1.52~m TCS telescope at Teide Observatory in Spain. MuSCAT2 is a sibling of MuSCAT, but has four channels for $g$-, $r$-, $i$- and $z_s$ bands. The CCD cameras of MuSCAT2 are identical to those of MuSCAT, but the pixel scale is 0\farcs44 pixel$^{-1}$, which provides a $7\farcm4\times 7\farcm4$ FOV. The observations were conducted with the exposure times of 3 to 60 sec depending on the band and sky condition. The obtained data were reduced in the same way as for the MuSCAT data. We adopted aperture radii of 8 -- 12 pixels (3\farcs5 -- 5\farcs2) depending on the band and night, which means that the companion star at 2\farcs5 away is contaminated into the photometric apertures in all bands.

\subsubsection{FTN 2m / MuSCAT3 Photometry}

We observed one transit of TOI-1685.01 on UT 2021 February 1 using the brand-new multiband imager MuSCAT3 \citep{2020SPIE11447E..5KN}, which was installed on the 2m Faulkes Telescope North (FTN) at Haleakala Observatory in Hawaii in late 2020. The telescope and instrument are operated by Las Cumbres Observatory. As with MuSCAT2, MuSCAT3 has four channels for $g$, $r$, $i$, and $z_s$ bands, but has wider format CCD cameras with a size of 2k $\times$ 2k. The pixel scale of each camera is 0\farcs266 pixel$^{-1}$, which provides a FOV of 9\farcm1 $\times$ 9\farcm1.  The observation was done with slightly out-of-focus and with the exposure times of 25, 9, 8, and 20~s for $g$, $r$, $i$, and $z_s$ bands, respectively. The obtained raw images were processed by the {\tt BANZAI} pipeline \citep{curtis_mccully_2018_1257560} for dark and flat corrections, and then aperture photometry was performed in the same way as for the MuSCAT and MuSCAT2 data. The adopted radii of photometric aperture were 14, 18, 14, and 16 pixels (3\farcs6, 4\farcs7, 3\farcs6, and 4\farcs2) for $g$, $r$, $i$, and $z_s$ bands, respectively.

\subsubsection{LCOGT Photometry}
We observed a full transit of TOI-1634.01 on UT 2020 September 30 in Pan-STARRS $z$-short band and a full transit of TOI-1685.01 on UT 2020 November 11 in Sloan $i'$ band from the Las Cumbres Observatory Global Telescope (LCOGT) \citep{Brown:2013} 1.0\,m network node at McDonald Observatory. We used the {\tt TESS Transit Finder}, which is a customized version of the {\tt Tapir} software package \citep{Jensen:2013}, to schedule our transit observations. The $4096\times4096$ LCOGT SINISTRO cameras have an image scale of $0\farcs389$ per pixel, resulting in a $26\arcmin\times26\arcmin$ field of view. The images were calibrated by the standard LCOGT {\tt BANZAI} pipeline \citep{McCully:2018}, and photometric data were extracted with {\tt AstroImageJ} \citep{Collins:2017}. The TOI-1634.01 observation was slightly defocused and used 40 second exposures and a photometric aperture radius of $5\farcs 8$ to extract the differential photometry, resulting in a photometric precision of $\sim 500$ ppm model residuals in 5 minute bins. The TOI-1685.01 observation was mildly defocused and used 50 second exposures and a photometric aperture radius of $7\farcs 8$ to extract the differential photometry, resulting in a photometric precision of $\sim 410$ ppm model residuals in 5 minute bins.

\subsubsection{OMM 1.6m / PESTO Photometry}
We observed a full transit of TOI-1685.01 at Observatoire du Mont-M\'{e}gantic, Canada, on UT 2020 March 8. The observations were made in the $i^\prime$ filter with a 15~s exposure time using the 1.6 m telescope of the observatory equipped with the 1024 $\times$ 1024 PESTO camera. PESTO has an image scale of $0\farcs466$ per pixel, which provides an on sky 7\farcm95 $\times$ 7\farcm95 FOV. The light curve extraction via differential photometry was accomplished using an aperture radius of $7\farcs0$ and {\tt AstroImageJ}. This software was also used for image calibration (bias subtraction and flat field division).

\subsection{High Resolution Imaging}
As part of the standard follow-up process, high resolution imaging was performed to search for blended bound and unbound stellar companions and account for their presence in the analysis \citep[e.g.,][]{ciardi2015}.  Observations were performed with the optical speckle camera 'Alopeke on Gemini-North for TOI-1634 and the near-infrared adaptive optics camera NIRC2 on Keck2 for TOI-1685.

\subsubsection{Gemini-North/'Alopeke Speckle Observations}

\begin{figure}
\centering
\includegraphics[width=8.5cm,pagebox=mediabox]{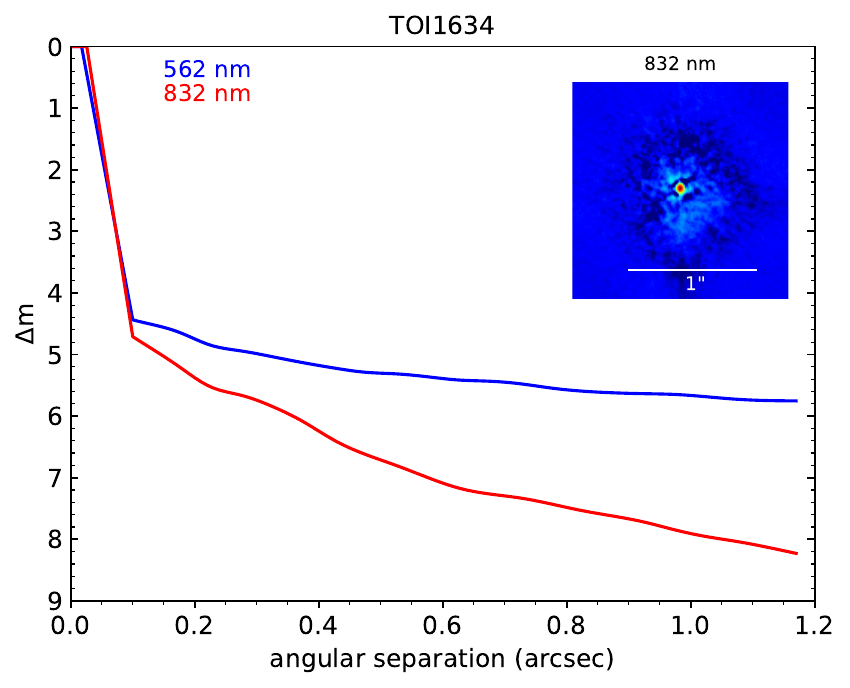}
\caption{
$5-\sigma$ contrast curves for TOI-1634 based on the Gemini-North/'Alopeke Speckle Observations. The inset displays the reconstructed image of the target. 
}
\label{fig:speckle}
\end{figure}

On UT December 2 2020, TOI-1634 was observed with the 'Alopeke speckle imager \citep{Scott2019}, mounted on the 8\,m Gemini North telescope on Mauna Kea. 'Alopeke simultaneously acquires data in two bands centered at 562\,nm and 832\,nm using high speed electron-multiplying CCDs (EMCCDs). We collected and reduced the data following the procedures described in \citet{Howell2011}. The resulting reconstructed image achieved a contrast of $\Delta\mathrm{mag}=8$ at a separation of 1\arcsec\ in the 832\,nm band (see Figure~\ref{fig:speckle}).
No secondary source was identified within $1\farcs2$ from TOI-1634. 

\subsubsection{Keck II/NIRC2 Observationa}
We observed TOI-1685 with near infrared (IR) high-resolution adaptive optics (AO) imaging at the Keck Observatory. 
We carried out the AO imaging 
using the NIRC2 instrument on Keck-II behind the natural guide star AO system. 
The observations were made on UT 2020 September 09 in the standard 3-point dither pattern that is used with NIRC2 to avoid the left lower quadrant of the detector, which is typically noisier than the other three quadrants. 
The dither pattern step size was set to $3\arcsec$ and was 
repeated twice, with each dither offset from the previous dither by $0\farcs5$.

\begin{figure}
\centering
\includegraphics[width=8.5cm,pagebox=mediabox]{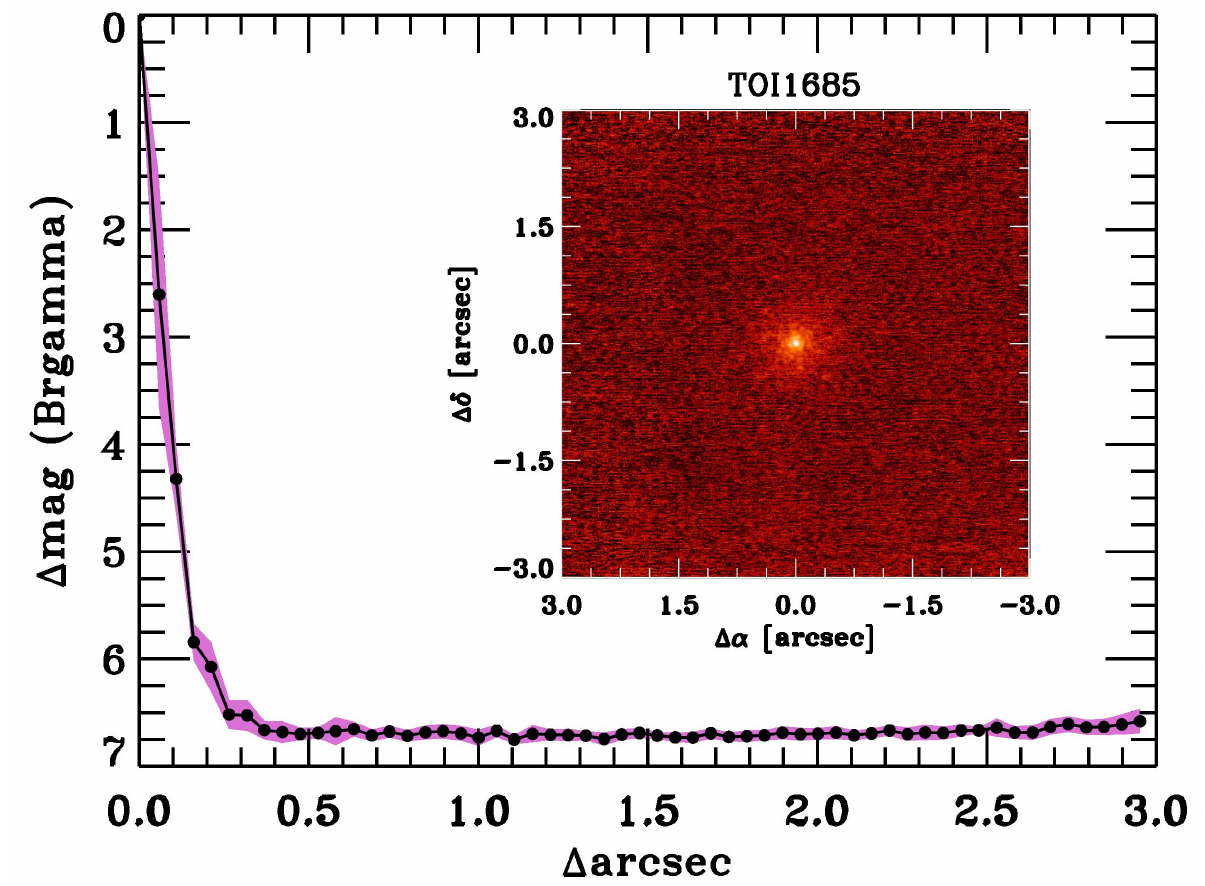}
\caption{Near IR AO image of TOI-1685 taken with NIRC2 on Keck2 and associated sensitivity curve.  The black points represent the $5-\sigma$ limits and are separated in steps of 1 FWHM ($\approx 0\farcs052$); the purple represents the azimuthal dispersion ($1\,\sigma$) of the contrast determinations (see text). The inset image is of the primary target showing no additional companions to within 3\arcsec\ of the target.  
}
\label{fig:1685ao_contrast}
\end{figure}

The observations were made in the narrow-band $Br-\gamma$ filter $(\lambda_o = 2.1686\,\mu\mathrm{m}; \Delta\lambda = 0.0326\,\mu$m) with an integration time of 1.5 seconds with one coadd per frame for a total of 13.5 seconds on target.  The camera was in the narrow-angle mode with a full FOV of $\approx 10\arcsec$ and a pixel scale of 
$\approx 0\farcs00994$ per pixel. 
The FWHM of the target in the combine image was $\approx 0\farcs052$, 
and no additional stellar companions were detected in the 
$6^{\prime\prime}\times 6^{\prime\prime}$ FOV (Figure \ref{fig:1685ao_contrast}).

The sensitivities of the final combined AO image were determined by injecting simulated sources azimuthally around the primary target every $20^\circ $ at separations of integer multiples of the central source's FWHM. 
Following e.g., \citet{2019AJ....158...79D}, we computed the $5\sigma$ 
sensitivity limit as a function of the radial distance from the target. 
The near IR AO sensitivity curve for TOI-1685 is shown in Figure \ref{fig:1685ao_contrast} along with an inset image zoomed to primary target showing no other companion stars.

\subsection{Spectroscopy} \label{sec:obs_sp}

\subsubsection{TRES Spectroscopy} \label{sec:tres}

We obtained reconnaissance spectra of TOI-1634 on UT 2020 February 1 and UT 2020 September 4 and of TOI-1685 on UT 2020 February 2 and UT 2020 February 3 using the Tillinghast Reflector Echelle Spectrograph (TRES; Furesz 2008) located at the Fred Lawrence Whipple Observatory (FLWO) in Arizona, USA. TRES has a resolving power of $\approx 44,000$ and a wavelength coverage of $385 - 910$ nm, 
and the spectra were extracted as described in \citet{2010ApJ...720.1118B}. 

RVs were determined from the TRES spectra using methods
outlined in \citet{2018AJ....155..125W}.  Briefly, molecular bands due
to TiO in the wavelength range 7065 - 7165 $\mathrm{\AA}$ found in aperture 41 of
the TRES spectra were cross-correlated with an observed template
spectrum of Barnard's Star (Gl 699).  We conducted a search for
maximum cross-correlation over a range of values of the rotational
broadening $v \sin i$ applied to the template spectrum prior to
correlation. As a result, we concluded there was no rotational broadening detectable
in either target and therefore fixed the rotational broadening to zero
for the final analysis.
There is a systematic uncertainty in the velocity zero point
of approximately 0.5 km s$^{-1}$ which may be important when considering the
absolute Barycentric RV, rather than the relative
velocity differences between the epochs.
We obtained $\mathrm{RV}=-17.066$ km s$^{-1}$ (2020 February 1) 
and $-17.105$ km s$^{-1}$ (2020 September 4) for TOI-1634, 
and $\mathrm{RV}=-43.306$ km s$^{-1}$ (2020 February 2) and $-43.219$ km s$^{-1}$
(2020 February 3) for TOI-1685. 
For each target, the two spectra were secured at near opposite quadratures 
in orbital phase based on the the TESS ephemerides. 
Therefore, the absence of large RV variations ($\gtrsim 0.5$ km s$^{-1}$) ruled out stellar and brown-dwarf companions as the source of the transits for both targets.

\subsubsection{Subaru/IRD Spectroscopy}

For precise RV measurements of TOI-1634 and TOI-1685, we carried out near 
IR observations of those two M dwarfs using Subaru/IRD between 2020 September 
and 2021 February under the Subaru IRD TESS intensive follow-up program (ID: S20B-088I). 
Every month during the period, we observed the two targets on $2-3$ different nights when the program 
was assigned. On some of those nights, we visited the target stars twice within a night 
(two visits separated by a few hours) in order to mitigate the impact of the $1-$day observing window, which happens to be close to the period of TOI-1634.01. 
IRD is a fiber-fed spectrograph placed in a temperature stabilized chamber, 
which can simultaneously cover broadband near IR wavelengths from 930 nm to 1740 nm 
with a spectral resolution of $\approx 70,000$ 
\citep{2012SPIE.8446E..1TT, 2018SPIE10702E..11K}. 
Stellar light collected by the telescope is first squeezed by the AO system on Subaru \citep{2008SPIE.7015E..10H}, which is then injected into the spectrograph through a multi-mode fiber.  
For TOI-1634, the companion star at $2\farcs5$  was resolved in IRD's fiber injection 
module camera, and we ensured that only the primary (brighter) star was injected into
the fiber. 
To trace the temporal instrumental stability, a secondary fiber is inserted into 
the spectrograph for the simultaneous wavelength calibration, to which the laser-frequency comb (LFC) is usually injected. 
The integration times for both targets were set to $720-1200$ sec for each exposure, depending on the observing condition. We also observed at least one telluric standard star (A0 or A1 star) on each night to correct for the telluric lines in extracting the template spectrum for the RV analysis. 

Raw IRD data were reduced by the standard procedure using IRAF \citep{1993ASPC...52..173T}
as well as our custom codes to process the detector's bias and wavelength calibrations by LFC spectra \citep{2018SPIE10702E..60K, 2020PASJ...72...93H}. 
The reduced one-dimensional spectra have a typical signal-to-noise (S/N) ratio of
$60-95$ per pixel at 1000 nm for both targets. Analyzing these reduced spectra, we extracted the RV for each frame. The RV analysis pipeline for IRD is described in \citet{2020PASJ...72...93H}; 
in short, individual observed spectra are first processed to create the stellar template spectrum, which is free from the telluric features and instrumental broadening. Using this stellar template as well as the instantaneous instrumental profile (IP) of the spectrograph (based on each LFC spectrum), each spectrum is fitted with the forward modeling technique. 
The typical RV internal errors are $3-4$ m s$^{-1}$ for both targets.

\section{Analyses and Results} \label{sec:ana}

\subsection{Estimation of Stellar Parameters} \label{sec:parameters}

In this subsection, we will estimate the stellar parameters based on three independent methods. We then derive the most reliable stellar parameters jointly using those estimations.

\subsubsection{Analysis of TRES spectra}

To estimate the basic stellar parameters, we independently analyzed the optical high-resolution spectra taken by TRES and near IR spectra by IRD. 
For the TRES spectra, we made use of {\tt SpecMatch-Emp} \citep{2017ApJ...836...77Y} to determine the effective temperature $T_\mathrm{eff}$, radius $R_\star$, and iron abundance [Fe/H] of the stars. 
The code attempts to fit an observed (input) high-resolution spectrum to a number of library spectra, whose stellar parameters were well determined, and find the best-matched stars in the library,  by which the 
stellar parameters for the input spectrum are determined by interpolations. 
{\tt SpecMatch-Emp} returned $T_\mathrm{eff}=3474\,\pm\,70$ K and $3468 \,\pm\,70 $ K, 
 $R_\star= 0.435\, \pm \,0.044\,R_\odot$ and $0.417\, \pm \,0.042\,R_\odot$, and 
 $\mathrm{[Fe/H]}=0.13\, \pm \,0.12$ dex and $0.03\, \pm \,0.12$ dex, 
 for TOI-1634 and TOI-1685, respectively. 

\subsubsection{Analysis of IRD spectra}

To estimate the atmospheric parameters for the two targets, 
we also analyzed the IRD spectra. Since many parts of the original IRD spectra suffer from significant telluric features (both absorptions and airglow emissions), we used the template spectra extracted for the RV analyses (Section \ref{sec:obs_sp}), in which telluric features were removed and multiple frames were combined. 
The template spectra were then subjected to the analysis tool developed by
\citet{2020PASJ...72..102I}. 
The analysis is based on a line-by-line comparison between the equivalent widths (EWs) from observed spectra and those from synthetic spectra.
The synthetic spectra were calculated with a one-dimensional LTE spectral synthesis code that is based on the same assumptions as the model atmosphere program of \citet{1978A&A....62...29T}.
For the atmospheric layer structure, we interpolated the grid of MARCS models \citep{2008A&A...486..951G}.
The surface gravity $\log{g}$ and micro-turbulent velocity were needed to be assumed for the analysis.
We referred to TIC for $\log{g}$ values calculated from masses and radii \citep{2019AJ....158..138S}, which were estimated from the mass-$M_K$ relation in \citet{2019ApJ...871...63M} and the radius-$M_K$ relation in \citet{2015ApJ...804...64M}, respectively.
The microturbulent velocity was fixed at $0.5 \pm 0.5$ km s$^{-1}$ for both objects for simplicity.

First, we used the FeH molecular lines in the Wing-Ford band at $990-1020$ nm for the $T_{\mathrm{eff}}$ estimation.
The band consists of more than 1,000 FeH lines, of which 57 lines with relatively clear line profiles were selected for the analysis. 
The adopted spectral line data are 
available from the MARCS web page\footnote{\url{https://marcs.astro.uu.se/}}.
We measured the EW of each FeH line by fitting the Gaussian profile and found $T_{\mathrm{eff}}$ at which the synthetic spectra best reproduce the EW by an iterative search.
Throughout this first step, we assumed the solar value for the metallicity.
The average of the $T_{\mathrm{eff}}$ estimates for each of the 57 lines was taken as the best estimate here. 
Its uncertainty was given as the line-to-line scatter calculated by the standard deviation over the estimates from all the lines.
Those procedures will be provided in more detail in  
Ishikawa et al. (2021, in preparation).  

As a second step, adopting the $T_{\mathrm{eff}}$ value estimated above, we determined the elemental abundances of Na, Mg, Si, Ca, Ti, Cr, Mn, and Fe from the corresponding atomic lines.
The details of the abundance analysis are given in \citet{2020PASJ...72..102I}, although they adopted literature values for $T_{\mathrm{eff}}$.
The spectral line data were taken from the Vienna Atomic Line Database (VALD; \citealt{1999A&AS..138..119K}, \citealt{2015PhyS...90e4005R}).
We selected the lines based on three criteria: (1) not suffering from blending of other absorption lines, (2) sensitive to elemental abundances, and (3) continuum level can be reasonably determined.
The EWs were measured by fitting synthetic spectra on a line-by-line basis.
We searched for an elemental abundance until the synthetic EW matches the observed one for each line and took the average for all the lines to estimate [X/H] for an element X.

Subsequently, we adopted the iron abundance [Fe/H] determined in the second step as the metallicity of the atmospheric model grid to redetermine the $T_{\mathrm{eff}}$ by the same procedure as in the first step. Then, we adopted the resulting $T_{\mathrm{eff}}$ to finally determine the elemental abundances including [Fe/H] again in the same way as in the second step.
The procedure up to this point allows the results of $T_{\mathrm{eff}}$ and abundances to converge well within the measurement errors.
Based on these analyses of IRD spectra, we obtained 
$T_\mathrm{eff}=3432\,\pm\,99$ K and $3428 \,\pm\,97 $ K, 
 $\mathrm{[Fe/H]}=0.27\, \pm \,0.12$ dex and $0.27\, \pm \,0.12$ dex
for TOI-1634 and TOI-1685, respectively. 
The abundances for the other elements are listed in Table \ref{hyo1}.

\subsubsection{Analysis of Broadband Photometry}\label{sec:SED}

\begin{figure}
    \centering
    \includegraphics[width=0.325\textwidth,trim=125 75 90 90,clip,angle=90]{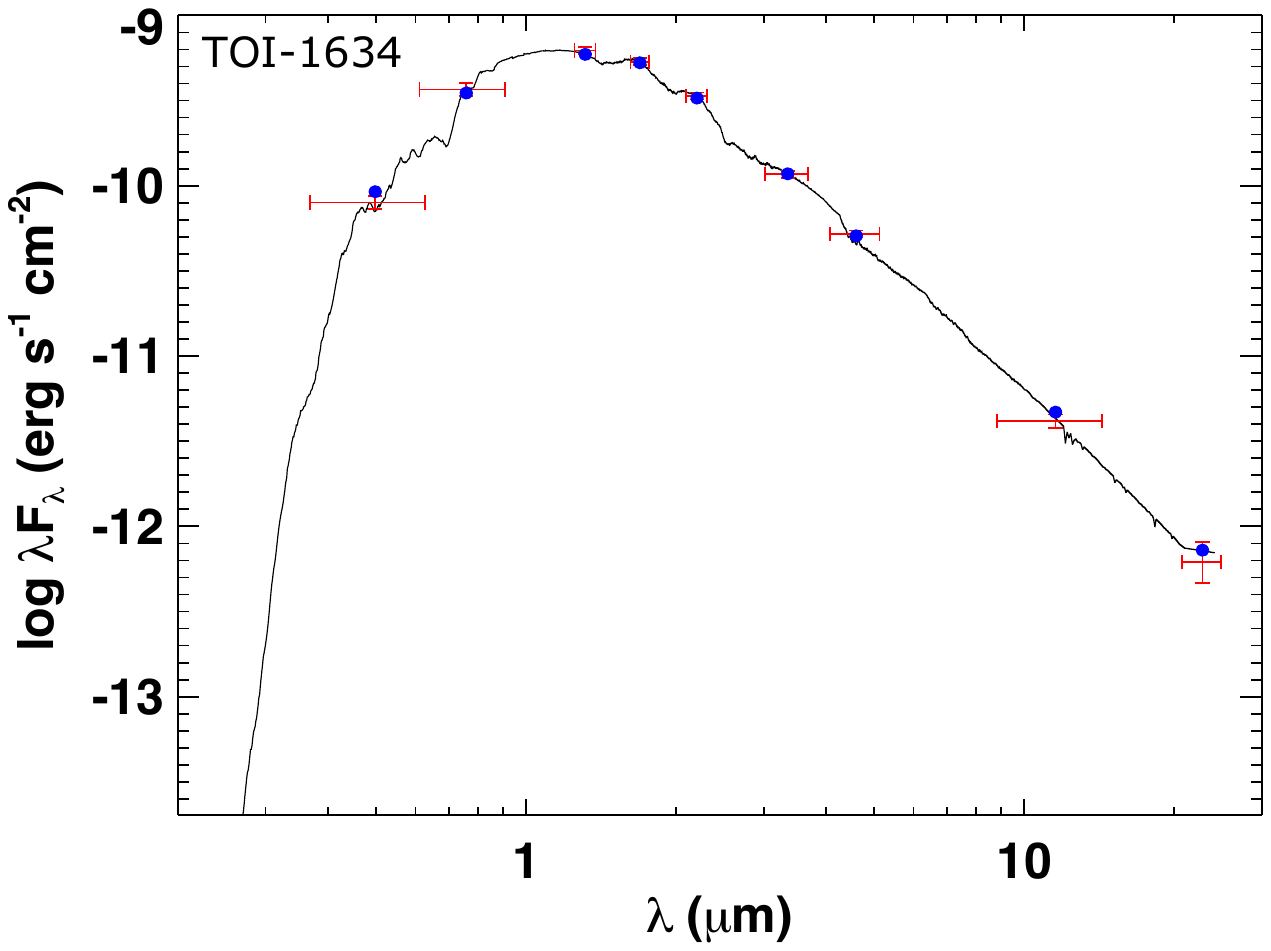}
    \includegraphics[width=0.365\textwidth,trim=75 75 90 90,clip,angle=90]{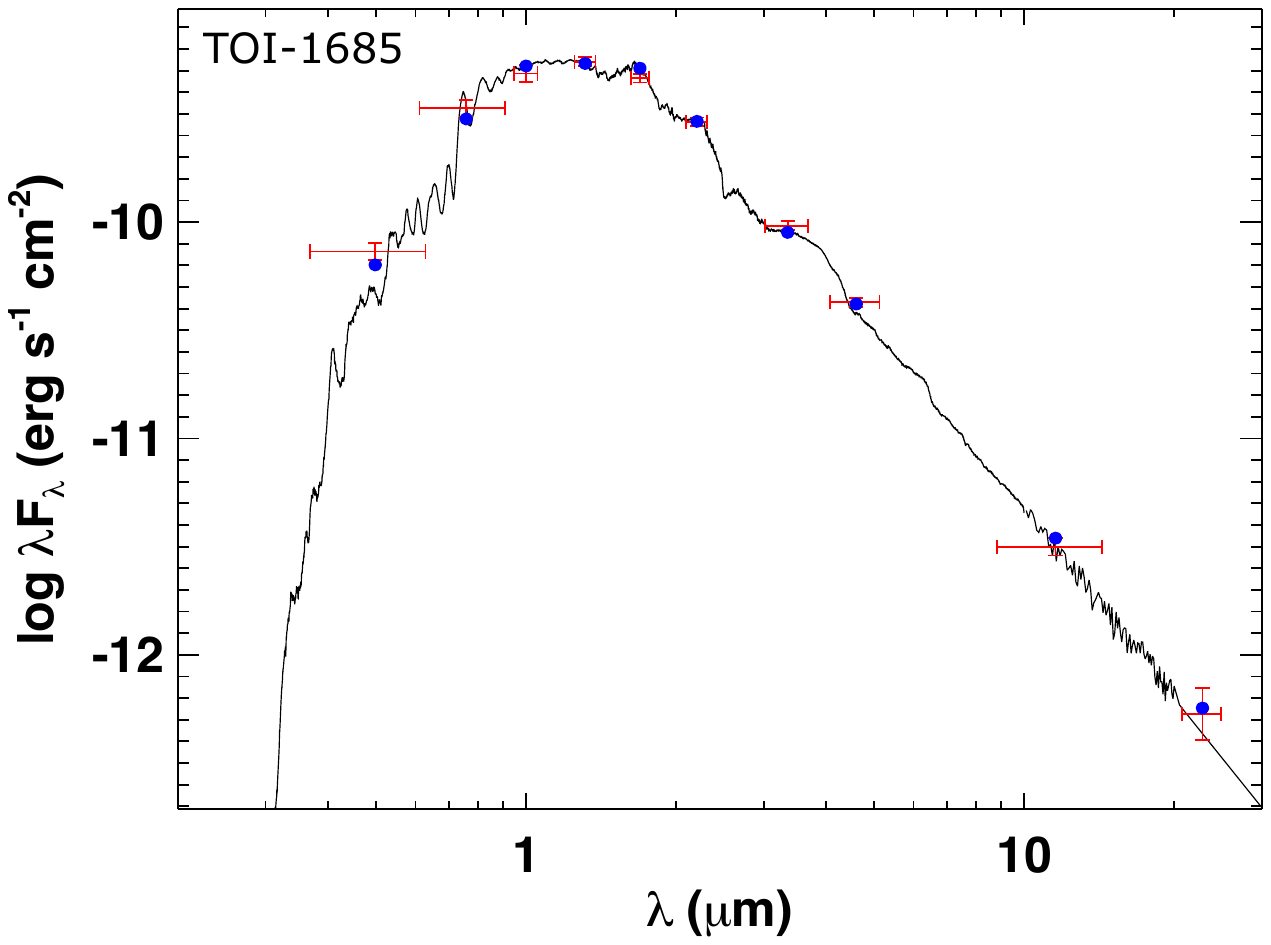}
\caption{Spectral energy distributions of TOI-1634 (top) and TOI-1685 (bottom). Red symbols represent the observed photometric measurements, where the horizontal bars represent the effective width of the passband. Blue symbols are the model fluxes from the best-fit NextGen atmosphere model (black).  \label{fig:sed}}
\end{figure}

We also performed an analysis of the broadband spectral energy distribution (SED) of the star together with the {\it Gaia\/} EDR3 parallax \citep[with no systematic offset applied; see, e.g.,][]{StassunTorres:2021}, in order to determine an empirical measurement of the stellar radius, following the procedures described in \citet{Stassun:2016,Stassun:2017,Stassun:2018}. We pulled the $JHK_S$ magnitudes from {\it 2MASS} \citep{2006AJ....131.1163S}, the W1--W4 magnitudes from {\it WISE} \citep{2010AJ....140.1868W}, the $G,\, G_{\rm BP},\, G_{\rm RP}$ magnitudes from {\it Gaia} \citep{2020arXiv201201533G}, and the $y$-band magnitudes from Pan-STARRS \citep{2020ApJS..251....7F}. Together, the available photometry spans the full stellar SED over the wavelength range 0.4--20~$\mu$m (see Figure~\ref{fig:sed}).  
We performed a fit using NextGen stellar atmosphere models, with $T_{\rm eff}$ and [Fe/H] as the free parameters; the extinction $A_V$ was fixed at zero due to the proximity of the stars. Integrating the (unreddened) model SEDs gives the bolometric flux at Earth, $F_{\rm bol}$. Finally, taking the $F_{\rm bol}$ and $T_{\rm eff}$ together with the {\it Gaia\/} parallax gives the stellar radius, $R_\star$. 
The SED analysis provided $T_\mathrm{eff}=3500\,\pm\,85$ K and $3475 \,\pm\,75 $ K, 
 $\mathrm{[Fe/H]}=0.0\, \pm \,0.5$ dex and $0.0\, \pm \,0.5$ dex, 
 $F_{\rm bol}=(7.64 \pm 0.27) \times 10^{-10}$ erg s$^{-1}$ cm$^{-2}$ and
 $(6.65 \pm 0.15) \times 10^{-10}$ erg s$^{-1}$ cm$^{-2}$, and
 $R_\star= 0.466\,\pm 0.024\,R_\odot$ and $0.473\,\pm\,0.021\,R_\odot$
 for TOI-1634 and TOI-1685, respectively.

\subsubsection{Joint Modeling of the Stellar Parameters} 

The three measurements (optical spectroscopy, near IR spectroscopy, 
and SED fitting) of $T_\mathrm{eff}$ and [Fe/H] yielded consistent results within their errors, and thus we computed 
the weighted means of those parameters to gain the final values (Table \ref{hyo1})
used in the subsequent analyses. 
Since these measurements ultimately rely on similar stellar atmosphere models
or the same calibration sources, we conservatively adopted the representative errors for the mean values of the two parameters (i.e., $70$ K for $T_\mathrm{eff}$ and $0.12$ dex for [Fe/H]). 
Based on the basic parameters derived above, we further estimated the other stellar parameters (i.e., the stellar mass $M_\star$, radius $R_\star$, surface gravity $\log g$, mean density $\rho_\star$, and luminosity $L_\star$), as well as refined the basic parameters (i.e., the stellar metallicity [Fe/H] and distance $d$) by combining all observed quantities in a consistent manner. 
In doing so, we took an approach described in \citet{2018AJ....155..127H}, but with the inclusion of Gaia parallaxes; 
since the observed quantities are redundant (e.g., there are two sets of estimates for the stellar radius) and can be correlated with each other through the empirical relations, 
we performed Markov Chain Monte Carlo (MCMC) simulations in which the $\chi^2$ statistic of the likelihood function ($\propto \exp(-\chi^2/2)$) is defined as
\begin{eqnarray}
\label{eq:1}
\chi^2 = \frac{(R_{\rm\star, TRES}-R_\star)^2}{\sigma_{R_\star, {\rm TRES}}^2} + 
\frac{(R_{\rm\star, SED}-R_\star)^2}{\sigma_{R_\star, {\rm SED}}^2} \nonumber\\
+ \frac{(m_{K_s, {\rm 2MASS}}-m_{K_s})^2}{\sigma_{m_{K_s}, {\rm 2MASS}}^2}, 
\end{eqnarray}
where $R_{\rm\star, TRES}$ and $R_{\rm\star, SED},$ are the stellar radii estimated by 
the optical spectroscopy and SED integration, and $\sigma_{R_\star, {\rm TRES}}$ and $\sigma_{R_\star, {\rm SED}}$ are their errors, respectively. 
The apparent $K_s-$band magnitude by 2MASS and its error are denoted by $m_{K_s, {\rm 2MASS}}$ and $\sigma_{m_{K_s}, {\rm 2MASS}}$, respectively.
The fitting parameters in the MCMC analysis are the absolute $K_s$ magnitude $M_{K_s}$, 
stellar metallicity $\mathrm{[Fe/H]}$, and the distance $d$ to the system. 
The modeled quantities $R_\star$ and $m_{K_s}$ in the right-hand side of Equation (\ref{eq:1})
are calculated from $M_{K_s}$, $\mathrm{[Fe/H]}$, and $d$ through the empirical relation by \citet{2015ApJ...804...64M} and $m_{K_s}-M_{K_s} = 5.0\log d - 5.0$. We assume $A_V=0$, 
given the proximity of the two stars to Earth. 
We imposed Gaussian priors on $\mathrm{[Fe/H]}$ and $d$ based on the weighted mean
value and its error for $\mathrm{[Fe/H]}$ derived above, and the Gaia parallax \citep{2020arXiv201201533G}. 
In implementing the MCMC analysis, we computed $M_\star$ via the empirical relation 
of \citet{2019ApJ...871...63M} from $M_{K_s}$ and [Fe/H], as well as the surface gravity $\log g$, the mean density $\rho_\star$, and the luminosity $L_\star$ for each step of the chain. 
For $L_\star$, we sampled the $T_\mathrm{eff}$ values with the Gaussian distribution based on the values in Table \ref{hyo1}.

TOI-1634 has a companion star at $2\farcs5$ away from the primary star, but we were unable to identify the companion star in the 2MASS catalog. 
We inspected the 2MASS image for TOI-1634, and found that the companion star
was buried in the point spread function of the primary star, whose FWHM was found to be $2\farcs7-2\farcs8$). 
This suggests that the $K_s$ magnitude listed in Table \ref{hyo1} may be contaminated by the companion star, 
and the true magnitude of the primary star could be slightly fainter. 
To roughly estimate its impact, we used the Dartmouth
isochrone model \citep{2008ApJS..178...89D} and inferred the mass of the companion. 
Since the Dartmouth isochrones list the Gaia magnitudes as a function of stellar mass for a given set of stellar age and metallicity, we employed the Gaia $G_{\rm RP}$ magnitude to constrain the companion's mass. The magnitude difference of $\Delta G_{\rm RP}=2.959$ between TOI-1634 and its companion
translates to the companion's mass
of $\approx 0.12\,M_\odot$ on the assumption that TOI 1634's mass is roughly $\approx 0.46\,M_\odot$. When those masses are adopted, the isochrones predict that the
magnitude difference in the $K_s$ band should be $\Delta m_{K_s}\approx 2.8-3.0$ mag, 
implying that the true $m_{K_s}$ of the primary star is $\approx 0.07$ mag fainter than the reported one. 
With this in mind, we adopted $m_{K_s}=8.67 \pm 0.07$ instead of $m_{K_s}=8.600 \pm 0.014$ for TOI-1634 (in addition to shifting the center value of the magnitude, 
we conservatively added the systematic error of $0.07$ in $m_{K_s}$ in quadrature) 
and ran the MCMC analysis.
For TOI-1685, we directly input the 2MASS $K_s$ magnitude 
in the code. MCMC simulations were implemented using our custom code
\citep[e.g.,][]{2015ApJ...799....9H} 
with the chain length of $10^6$ after the burn-in chains. 
The final derived parameters based on this MCMC analysis ($d$, [Fe/H], $\log g$, $M_\star$, $R_\star$, $\rho_\star$, and $L_\star$) are summarized 
in Table \ref{hyo1}.

Using the Gaia EDR3 information as well as the absolute RVs from the TRES spectra, 
we also computed the Galactic space velocities ($U,\,V,\,W$) for the two stars
with respect to the Sun (Table \ref{hyo1}). 
The low space velocities for both targets indicate those stars belong to the
thin disk. 
Velocity dispersions in the Galactic coordinate system are generally correlated 
with stellar age. 
Following the methodology described in \citet{2017ApJ...845..110B}, we computed the posterior distributions for the ages of the two stars. In doing so, we adopted the prescription given by \citet{2015MNRAS.449.3479S} for the velocity-dispersion evolution of the thin-disk stars with the Sun's peculiar velocity from \citet{2016ARA&A..54..529B}, and we used two different age priors: a uniform prior ($0<\mathrm{age}\leq 14$ Gyr) and the age probability distribution in the Geneva-Copenhagen Survey (GCS) catalog \citep{2011A&A...530A.138C}. 
Based on the age posterior distributions, we found TOI-1634 has the age of $3.2_{-2.8}^{+6.2}$ Gyrs (uniform prior) and $5.2_{-2.8}^{+4.0}$ Gyrs (GCS prior)
and that of TOI-1685 is $5.0_{-3.7}^{+5.6}$ Gyrs (uniform prior) and $5.7_{-3.0}^{+3.8}$ Gyrs (GCS prior), respectively. These results suggest the UVW velocities are not useful for constraining the ages of the two targets. 
We also confirmed that nether of the targets belong to nearby young associations
based on the \texttt{BANYAN $\Sigma$} tool \citep{2018ApJ...856...23G}.

\subsection{Analysis of Transit Light Curves} \label{sec:photometry}

\begin{figure}
    \centering
    \includegraphics[width=0.45\textwidth]{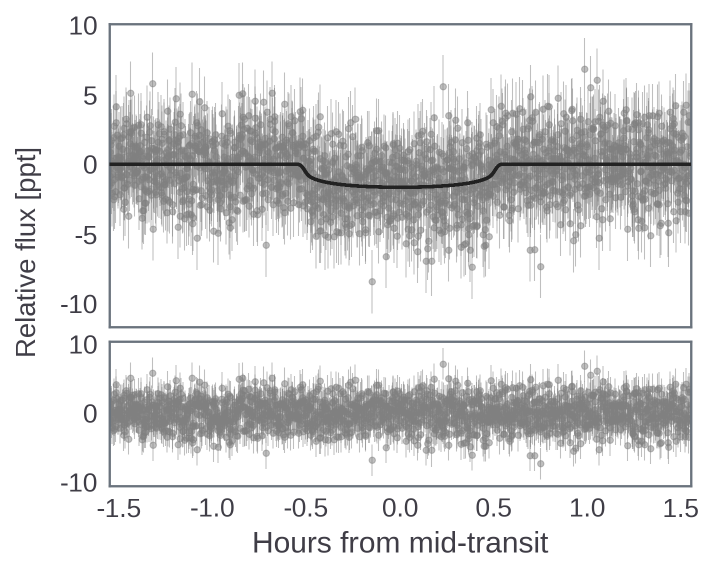}
    \caption{Phase-folded TESS photometry with transit model for TOI-1634.01 (top), and the residuals from the fit (bottom).}
    \label{fig:toi1634-tess}
\end{figure}

\begin{figure}
    \centering
    \includegraphics[width=0.45\textwidth]{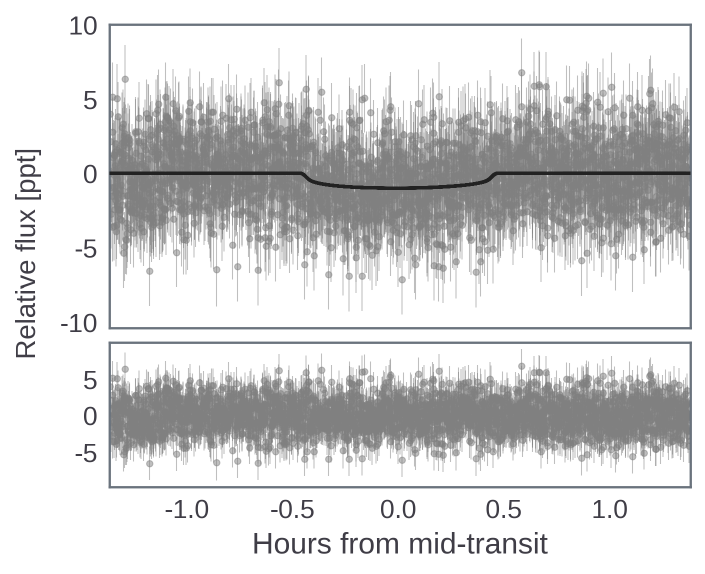}
    \caption{Same as Figure~\ref{fig:toi1634-tess} but for TOI-1685.01.}
    \label{fig:toi1685-tess}
\end{figure}

We fit the TESS, MuSCAT, MuSCAT2, MuSCAT3, OMM, and LCO datasets using the \texttt{PyMC3} \citep{pymc3}, \texttt{exoplanet}\footnote{\url{https://docs.exoplanet.codes/en/stable/}} \citep{exoplanet}, \texttt{starry} \citep{luger18}, \texttt{celerite2} \citep{celerite1,celerite2} software packages. To account for systematics in the ground-based datasets we included a linear model of the covariates: airmass, pixel centroids, and the pixel response function peak and width. In addition, we included a Gaussian Process \citep[GP][]{RasmussenWilliams2005} model to account for residual correlated noise not accounted for by the linear model, using a Mat\'ern-3/2 covariance function. The transit model parameters we fit were: stellar mass and radius, quadratic limb darkening parameters (two per bandpass), orbital period ($P$), time of transit center ($T_c$), planet to star radius ratio ($R_p/R_\star$), and impact parameter ($b$). We assumed a circular orbit and placed Gaussian priors on the stellar mass and radius based on the results in Table~\ref{hyo1}. We also placed Gaussian priors on the limb darkening coefficients based on interpolation of the parameters tabulated by \citet{Claret2012,Claret2017}, propagating the uncertainties in the stellar parameters in Table~\ref{hyo1} via Monte Carlo simulations. 

\begin{figure*}
    \centering
    \includegraphics[width=0.9\textwidth]{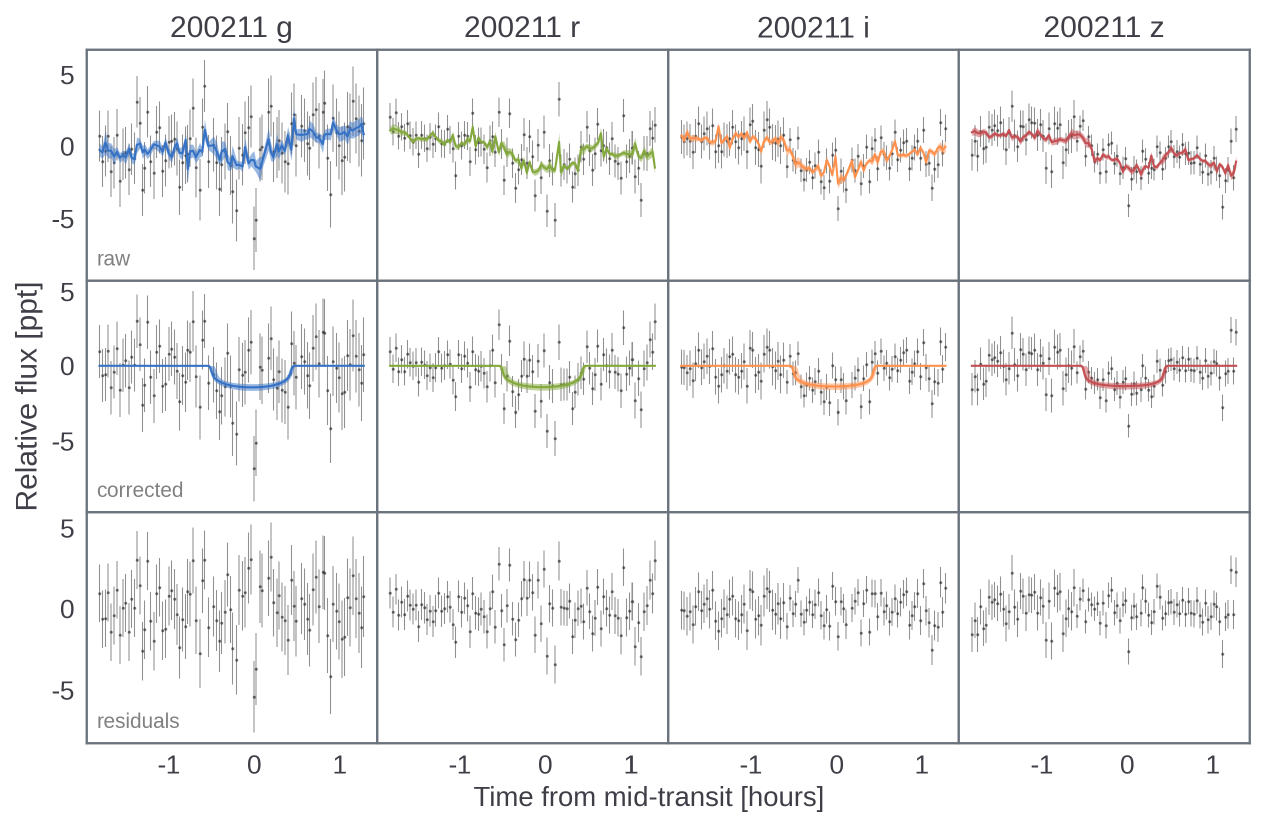}
    \caption{MuSCAT2 photometry of TOI-1634.01 taken on UT 2020 February 11. The upper row shows the raw photometry with full systematics and transit model in each bandpass, the middle row shows the systematics-corrected photometry with only the transit model, and the bottom row shows the residuals from the fit.}
    \label{fig:toi1634-m2}
\end{figure*}

\begin{figure*}
    \centering
    \includegraphics[width=0.9\textwidth]{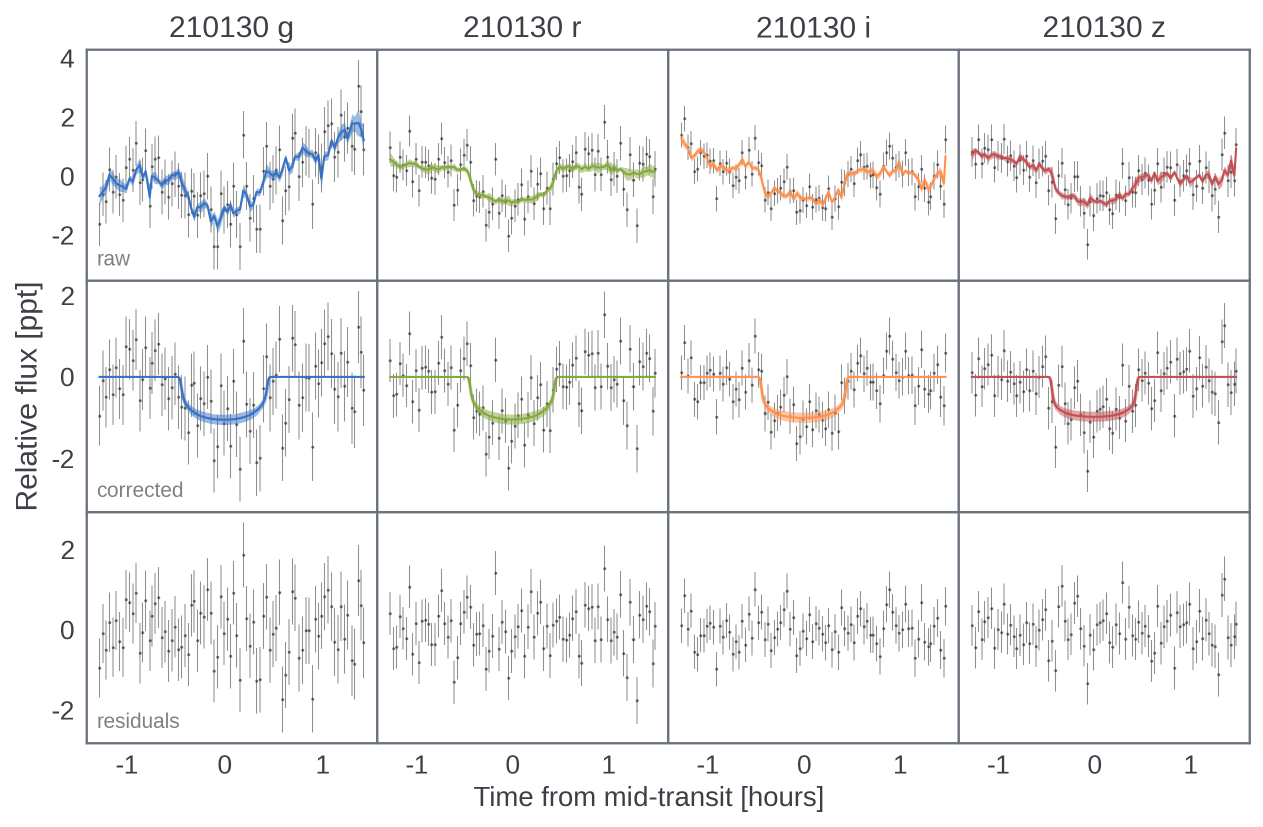}
    \caption{Same as Figure~\ref{fig:toi1634-m2}, but for the MuSCAT3 photometry of TOI-1685.01 taken on UT 2021 January 30.}
    \label{fig:toi1685-m3}
\end{figure*}

\begin{figure}
    \centering
    \includegraphics[width=0.5\textwidth]{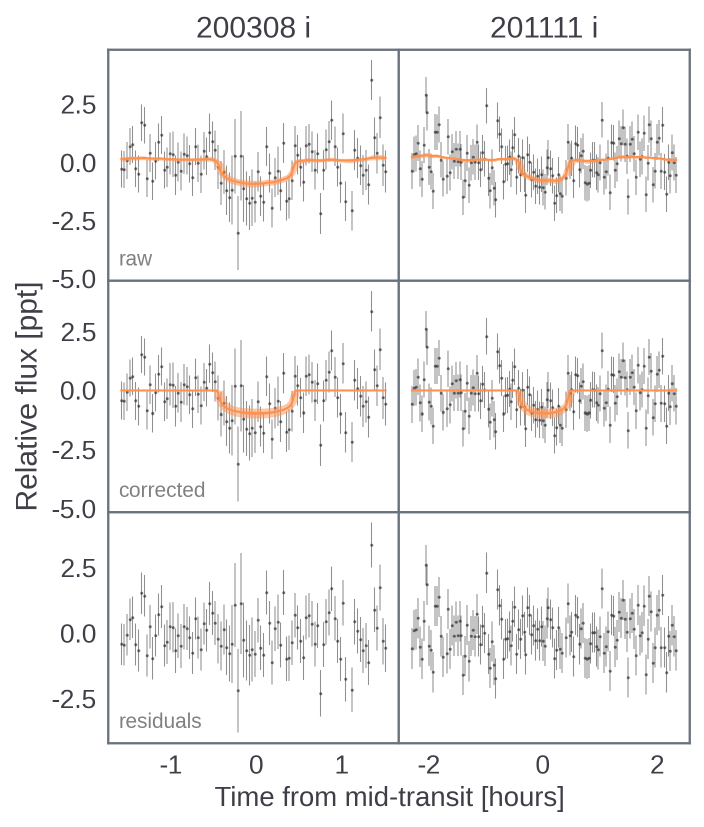}
    \caption{Same as Figure~\ref{fig:toi1634-m2}, but for the OMM (left) and LCO (right) photometry of TOI-1685.01 taken on UT 2020 March 8 and November 11, respectively.}
    \label{fig:toi1685-omm-lco}
\end{figure}

We used the gradient-based {\tt BFGS} algorithm \citep{NoceWrig06} implemented in {\tt scipy.optimize} to find initial maximum a posteriori (MAP) parameter estimates. We used these estimates to initialize an exploration of parameter space via ``no U-turn sampling'' \citep[NUTS,][]{HoffmanGelman2014}, an efficient gradient-based Hamiltonian Monte Carlo (HMC) sampler implemented in {\tt PyMC3}.
We first conducted a fit to the TESS data using a window centered on each transit of width three times the full transit duration ($3 \times T_{14}$), including a local linear time baseline function for each window to account for stellar variability. The folded TESS data and best fit transit models are shown in Figures \ref{fig:toi1634-tess} and \ref{fig:toi1685-tess}. We then fit each of the ground-based transit datasets using Gaussian priors derived from the impact parameter and orbital period posteriors of the TESS fit, in addition to the stellar mass, radius, and limb darkening priors. We assumed an achromatic transit model, and shared the GP hyperparameters between photometric bands taken simultaneously by MuSCAT1/2/3. Examples of the ground-based data and model fits for the various instruments used in this work are shown in Figures \ref{fig:toi1634-m2}, \ref{fig:toi1685-m3}, and \ref{fig:toi1685-omm-lco}. 
Due to the increased photometric scatter of the target stars in bluer bandpasses, we performed tests to determine whether the precision of our ground-based simultaneous multi-band transit measurements could be improved by using only the redder bandpasses. Despite the relatively low SNR of the transit signal in $g$ band, for the dataset shown in Figure~\ref{fig:toi1634-m2}, we found that excluding $g$ band from the fit (i.e. using only $r$, $i$, and $z_s$ bands) resulted in 18\% worse precision in $T_c$, and 11\% worse precision in $R_p/R_\star$. Similarly, we found that excluding both $g$ and $r$ bands from the fit resulted in 55\% worse precision in $T_c$ and 43\% worse precision in $R_p/R_\star$. We thus opted to include all bands in our fits in order to take advantage of the maximum precision afforded by our datasets.
Finally, we computed a weighted mean of the measurements of $R_p/R_\star$ from each dataset, and used the individual transit time posteriors to compute a linear orbital ephemeris and search for transit timing variations; the resulting parameter estimates are listed in Table~\ref{hyo2}. 

\begin{table*}[t]
\centering
\caption{Planetary Parameters of TOI-1634b and TOI-1685b}\label{hyo2}
\begin{tabular}{lcc}
\hline\hline
Parameter & TOI-1634b & TOI-1685b \\\hline
\multicolumn{3}{l}{\it Transit parameters}  \\
$P$ (days) & $0.9893436 \pm 0.0000020$ & $0.6691416 \pm 0.0000019$ \\
$T_c$ (BJD-2457000) & $1791.51495 \pm 0.00053$ & $1816.2255 \pm 0.0011$ \\
$b$ & $0.375 \pm 0.049$ & $0.416 \pm 0.053$ \\
$R_p/R_\star$ & $0.0356 \pm 0.0010$ & $0.0291 \pm 0.0010$ \\ 
\hline
\multicolumn{3}{l}{\it Derived parameters}  \\
$R_p$ ($R_\oplus$) & $1.749 \pm 0.079$ & $1.459 \pm 0.065$ \\
$M_p$ ($M_\oplus$) & $10.14 \pm 0.95$ & $3.43 \pm 0.93$\\
$\rho_p$ (g cm$^{-3}$) & $10.4_{-1.6}^{+1.9}$ & $6.1_{-1.7}^{+1.9}$ \\
$a$ (au) & $0.01490 \pm 0.00017$ & $0.011557 \pm 0.000092$ \\
$i_o$ (deg) & $86.98 \pm 0.41$ & $85.59 \pm 0.58$ \\
$T_\mathrm{eq}$ ($A_B=0$) (K) & $920 \pm 25$ & $1052 \pm 26$\\
$T_\mathrm{eq}$ ($A_B=0.3$) (K) & $842 \pm 23$ & $962 \pm 24$ \\
\hline
\end{tabular}
\end{table*}

\subsection{Rotation Analysis} \label{sec:photometry-rotation}

\begin{figure}
\centering
\includegraphics[width=8.5cm]{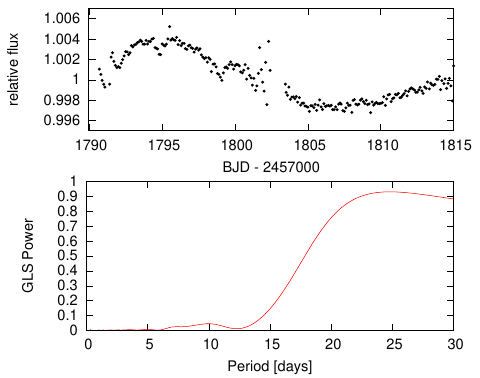}
\caption{
PLD-corrected binned TESS light curve for TOI-1634 (upper panel) and its GLS periodogram (bottom panel). 
}
\label{fig:period1}
\end{figure}
\begin{figure}
\centering
\includegraphics[width=8.5cm]{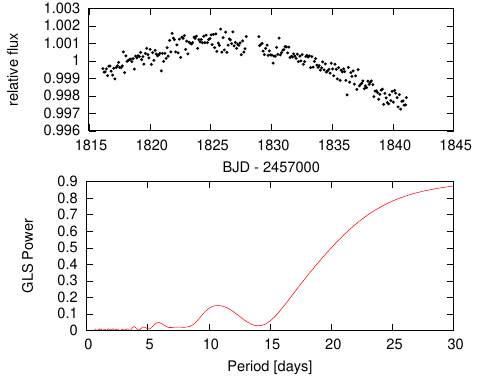}
\caption{
PLD-corrected binned TESS light curve for TOI-1685 (upper panel) and its GLS periodogram (bottom panel).
}
\label{fig:period2}
\end{figure}
As a last piece of the light curve analysis, we performed a periodogram analysis 
on the TESS
light curves for both targets to search for possible rotational modulations.  
The rotation period is one of the basic parameters to characterize the host star, which is also useful to disentangle the real planetary signal from the stellar activity in modeling the observed RV variations \citep[e.g.,][]{2015ApJ...808..127G, 2019MNRAS.490..698B}. 
We calculated the Generalized Lomb-Scargle (GLS) periodograms \citep{2009A&A...496..577Z} for the TESS light curves of TOI-1634 and TOI-1685 corrected for systematics using Pixel Level Decorrelation \citep[PLD;][]{Deming2015}, as implemented in the {\tt lightkurve} package \citep{lightkurve}. The SPOC pipeline removes instrumental correlated noise from the TESS light curves, but it can also remove astrophysical signals; we opt to use PLD instead, as it can correct systematics while preserving signals of interest, such as star spot modulation.
Figures \ref{fig:period1} and \ref{fig:period2} show the PLD light curves after binning (1 bin = 0.1 day) as well as the GLS periodograms for TOI-1634 and TOI-1685, respectively. 
Both light curves exhibit low-frequency modulations likely induced by surface spots, 
but in both cases the periodicity is ambiguous due to the short observing windows. 
The period of TOI-1634 could be around 24.8 days based on the GLS peak and 
visual inspection, but it may correspond to a multiple of the true rotation frequency. 
For TOI-1685, the light curve and periodogram indicate the rotation period of 
the star is much longer than the observing window (i.e., $P_{\rm rot}\gtrsim 30$ days). 

We also inspected the photometric data by the All-Sky Automated Survey for Supernovae 
\citep[ASAS-SN:][]{2014ApJ...788...48S, 2017PASP..129j4502K}, which recorded the magnitudes of target stars for more than five years. However, both GLS periodograms for TOI-1634 and TOI-1685 show no meaningful 
peak ($\mathrm{FAP}<1.0\,\%$), likely due to the low photometric precision ($\approx 1.5-2.0\,\%$) compared to the variability amplitude by stellar rotation 
\citep[typically less than 0.01 mag:][]{2016ApJ...821...93N, 2020ApJ...905..107M}. 
Unfortunately, available photometric data did not allow us to pin down the accurate
rotation periods for TOI-1634 and TOI-1685, but we confirmed that both targets are slowly rotating stars with $P_{\rm rot}\gtrsim 25$ days from the TESS light curves. 
This lower limit on $P_{\rm rot}$ corresponds to an upper limit of 
$\approx 0.90-0.92$ km s$^{-1}$ on $v\sin i$ for both stars.

The slow rotation of the two targets indicates that they are relatively old 
M dwarfs. The old ages are also corroborated by the lack of an emission line
in the chromospheric activity indicators. For instance, we inspected the H-$\alpha$ line in the TRES optical spectra for both targets, and found that they have the H-$\alpha$ ``absorption" line with no sign of emission at the line core. Such an absorption feature at H-$\alpha$ for an M3 dwarf implies that the stellar age is likely older than a few Gyr \citep[see e.g., Figure 6 of][]{2021arXiv210401232K} and the star has a long rotation period \citep[e.g.,][]{2017ApJ...834...85N}. 
This is also consistent with the lack of flares in the TESS light curves, whose rate provides a good indicator for the stellar age of mid-to-late M dwarfs \citep{2020ApJ...905..107M}.


\subsection{Period Analyses and Orbital Fits} \label{sec:RV}

In this subsection, we describe the period analyses and orbital fits to the RV data obtained by Subaru/IRD. 

\subsubsection{TOI-1634}

\begin{figure*}
\centering
\includegraphics[width=15cm]{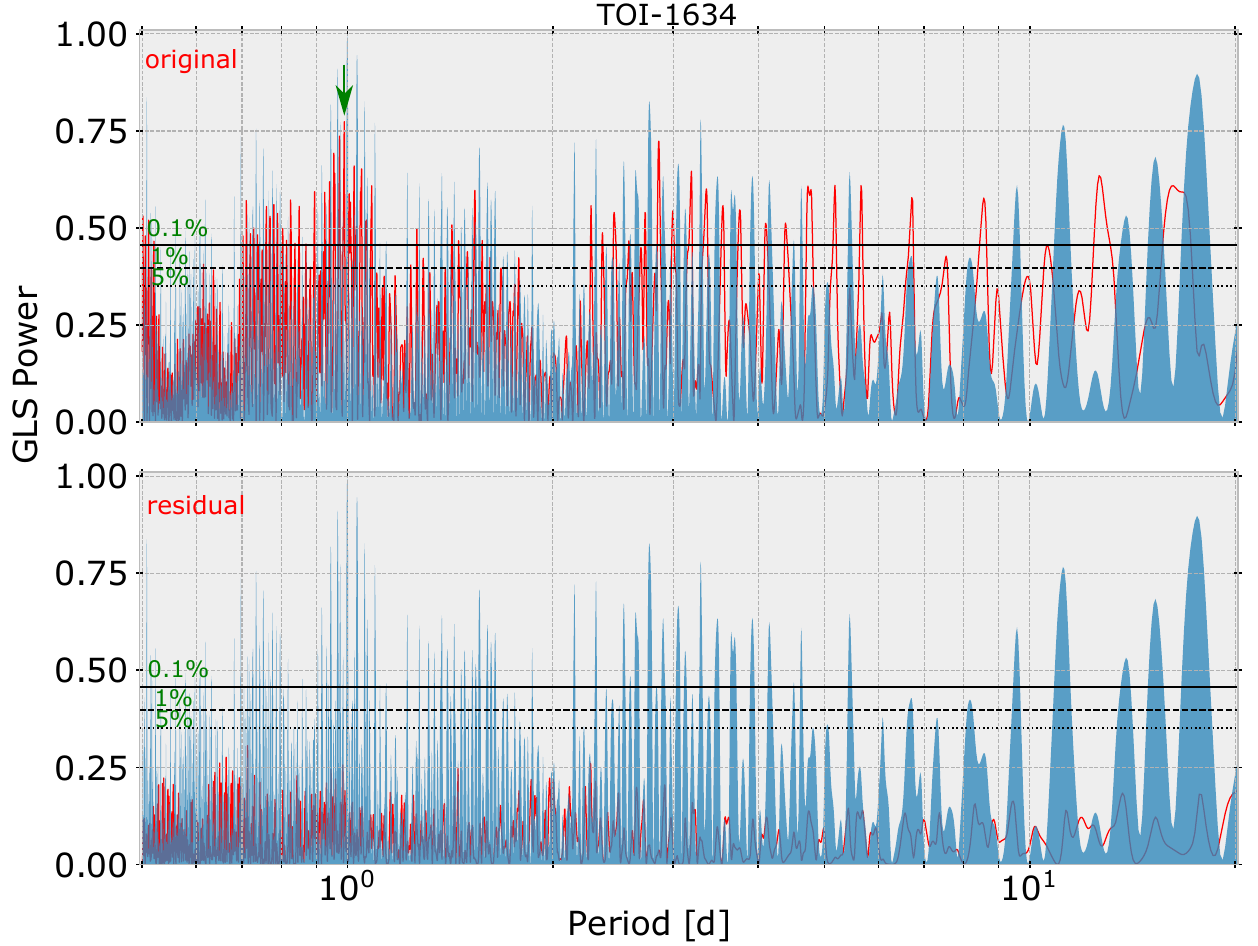}
\caption{
GLS periodograms for TOI-1634. 
The upper panel displays the periodogram in red on the original RV data. 
The GLS periodogram computed for the residual RV data after subtracting 
the best-fit Keplerian motion by the USP planet (TOI-1634b) is shown 
in the lower panel. In both panels, the window functions are shown by the blue
shaded regions. The highest peak in the upper panel, indicated by the green arrow,
precisely matches the correct period of TOI-1634b ($P=0.989$ day). 
}
\label{fig:rv-period}
\end{figure*}

The planetary transit was securely detected in the light curves by the ground-based
photometry (Figure \ref{fig:toi1634-m2}), in which the observed transit depths were consistent with the TESS photometry. 
However, the companion star at $2\farcs5$ was inside the 
photometric aperture\footnote{Since we defocused the images to achieve a better photometric precision, we are unable to distinguish the fluxes from the two stars.}, meaning that the ground-based photometry alone was
not capable of ruling out the possibility that the transits are originating from the
companion star (companion's flux contamination is larger than the transit depth). 
In order to check if our RV data alone indicates the presence of the USP planet around
TOI-1634, we performed the period analysis using the GLS tool \citep{2009A&A...496..577Z}
applied to the observed IRD-RV data. 
The upper panel of Figure \ref{fig:rv-period} shows the GLS periodogram for TOI-1634's raw RV data. There are multiple peaks with very low FAP's ($<0.1\,\%$) , but the highest peak shows up at the correct period of the transiting planet ($P=0.989$ day), which does not fall on the peaks of the window function (blue shaded area). 
Therefore, our RV data indicate additional, independent evidence of the USP planet
orbiting TOI-1634 and not orbiting its companion star.

Next, we attempted the orbital fit to the observed RVs. In doing so, we first estimated the impact of the companion star around TOI-1634; given the proximity to the star, the companion star at $2\farcs5$ away might have a non-negligible impact on the long-term RV baseline. 
With the distance of $d=35$ pc for TOI-1634, the angular separation of $2\farcs5$ 
translates to the projected separation of $88$ au, which approximately 
sets the lower limit to the semi-major axis of the binary orbit except for a highly eccentric orbit
(i.e., $a_{\rm binary}\gtrsim 88$ au). 
The RV acceleration of the primary star ($\dot{\gamma}$) around the center of mass 
of the system is expressed as
\begin{eqnarray}
\dot{\gamma} = \frac{GM_{\rm comp}}{a_{\rm binary}^2} \sin i_o 
\sin (f+\omega) \left( \frac{1+e\cos f}{1-e^2} \right)^2,
\end{eqnarray}
where $G$ is the gravitational constant, $M_{\rm comp}$ is the companion star's mass, 
$i_o$ is the orbital inclination, $f$ is the true anomaly, $e$ is the orbital eccentricity,
and $\omega$ is the argument of periastron. 
When we assume the companion's mass of $\approx 0.1\,M_\odot$ 
(see Section \ref{sec:parameters}) and $e=0$ for the
binary orbit, the lower limit on $a_{\rm binary}$ gives the maximum RV
acceleration as
\begin{eqnarray}
\label{eq:3}
|\dot{\gamma}| \leq \frac{GM_{\rm comp}}{a_{\rm binary}^2} \lesssim 7.3\times 10^{-3} \mathrm{~m~s^{-1}~day^{-1}}. 
\end{eqnarray}
In the presence of a moderate eccentricity, the upper limit of $\dot{\gamma}$ could be
a few times larger than the above value, depending on the orbital phase. 
Hence, this order-of-magnitude estimation suggests that the stellar companion may lead to an RV drift of up to a few m s$^{-1}$ over the course of $\approx 5$ months.

We constrained the visual binary orbital parameters using the \texttt{LOFTI\_gaiaDR2} software package \citep{2020ApJ...894..115P}. \texttt{LOFTI\_gaiaDR2} uses the instantaneous positions, proper motions, and masses of the components of visual binary stars to estimate their orbital parameters. We used the astrometric parameters from \textit{Gaia} EDR3 for this calculation, along with the stellar masses for the primary and secondary stars estimated in Section \ref{sec:parameters}. The \texttt{LOFTI\_gaiaDR2} posterior probability distribution has a slight preference for highly eccentric solutions (68\% confidence interval between $e= 0.61$ and $0.98$), but remains consistent with circular orbits. We note that these parameters should be taken with some skepticism because the astrometric solution for the secondary star shows excess scatter (with a Renormalized Unit Weight Error, or RUWE, of 1.7) which can indicate that it is itself an unresolved binary companion which can significantly affect its proper motion. Regardless, we conclude that the \textit{Gaia} positions and proper motions are not inconsistent with an eccentric visual binary orbit.

Based on these speculations, we modeled the observed RVs of TOI-1634 by the following equation, in which we allow for the presence of a possible RV trend:
\begin{eqnarray}
\mathrm{RV}(t) = K [\cos(f+\omega)+e\cos\omega] + \gamma + \dot{\gamma} (t-t_0),
\end{eqnarray}
where $K$ is the RV semi-amplitude and $\gamma$ is the RV offset of our data set. 
The time $t_0$ is an arbitrary origin of time, for which we 
adopt the time of the first RV point in the whole data set. 
We optimized the orbital parameters ($K$, $e\cos\omega$, $e\sin\omega$, $\gamma$, $\dot{\gamma}$)
using MCMC \citep{2015ApJ...799....9H} with uniform priors for all parameters. 
In the fit, we fixed $P$ and $T_c$ based on the transit ephemeris (Table \ref{hyo2}).

\begin{table*}[t]
\centering
\caption{Results of the Orbital Fits}\label{hyo3}
\begin{tabular}{l|ccc|ccc}
\hline\hline
 & \multicolumn{3}{c|}{TOI-1634b} & \multicolumn{3}{c}{TOI-1685b} \\
Parameter & with $\dot{\gamma}$ ($e=0$)$^\star$ & with $\dot{\gamma}$ ($e\neq0$) & no $\dot{\gamma}$ ($e=0$) & with Feb-02 ($e=0$) & no Feb-02 ($e=0$)$^\star$ & no Feb-02 ($e\neq0$) \\\hline
$K$ (m s$^{-1}$) & $11.1\pm 1.0$ & $11.31\pm 0.99$ & $11.80\pm 0.91$ & $3.3\pm 1.1$ & $4.2\pm 1.1$ & $7.0_{-1.6}^{+1.5}$ \\
$e\cos\omega$ & 0 (fixed) & $-0.118_{-0.038}^{+0.040}$ & 0 (fixed) & 0 (fixed) & 0 (fixed) & $0.278 \pm 0.076$ \\
$e\sin\omega$ & 0 (fixed) & $0.010_{-0.077}^{+0.075}$ & 0 (fixed) & 0 (fixed) & 0 (fixed) & $0.03_{-0.21}^{+0.20}$ \\
$\dot{\gamma}$ (m s$^{-1}$ day$^{-1}$) & $-0.023\pm 0.012$ & $-0.023\pm 0.012$ & 0 (fixed) & 0 (fixed) & 0 (fixed) & 0 (fixed) \\
BIC  & 70.6 & 69.1 & 70.0 & 54.4 & 43.1 & 39.0 \\
\hline
\end{tabular}
Note: For each planet, the fitting result adopted to compute the planet mass is indicated by $\star$.
\end{table*}

\begin{figure*}
\centering
\includegraphics[width=17.5cm]{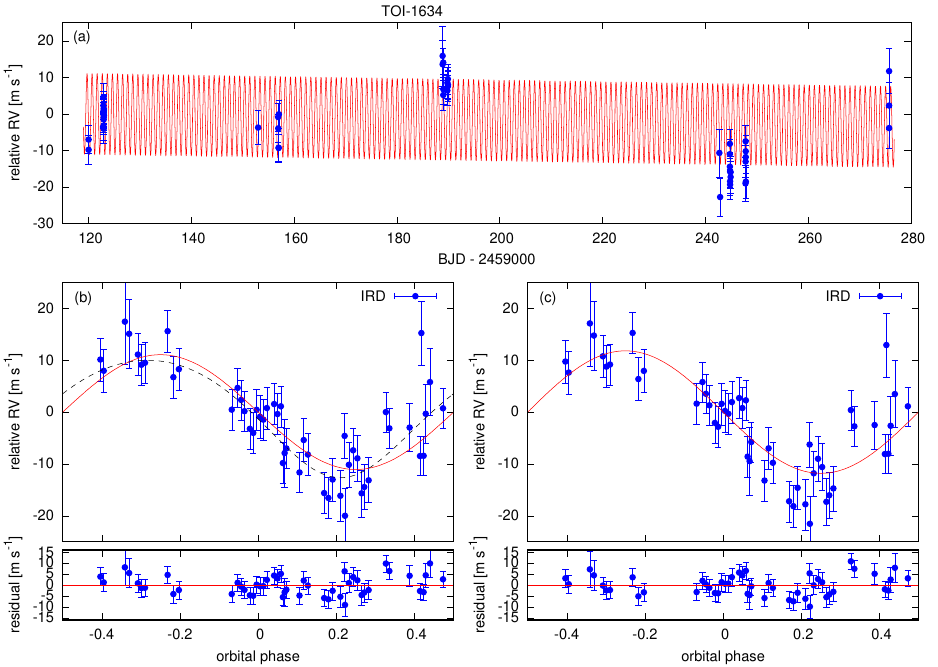}
\caption{
Observed RV variations for TOI-1634. (a) The original RVs are plotted as a function of
BJD, along with the best-fit model including a linear RV trend ($e=0$). 
(b) Phase-folded RV curve for TOI-1634b after subtracting the linear RV trend. 
The best-fit models with the circular and eccentric orbits are drawn by 
solid (red) and dashed curves, respectively. 
(c) Phase-folded RV curve assuming no RV trend is present ($\dot{\gamma}=0$) in the data. The best-fit model for $e=0$ is shown in red.
}
\label{fig:toi1634_rv}
\end{figure*}

We attempted the orbital fits assuming both circular and eccentric orbits. 
The results of those fits are shown in Table \ref{hyo3} (``with $\dot{\gamma}$" columns). 
To discuss the significance of the non-zero eccentricity, we compared the Bayesian Information Criterion (BIC), 
which is computed by $\mathrm{BIC} = \chi_{\rm best}^2 +k \ln (N)$, where $k$ is the number of fitting parameters
and $N$ is the number of data points. 
Comparing the two BIC values for the above solutions, we found 
$\Delta\mathrm{BIC}= \mathrm{BIC}_{\rm e=0}-\mathrm{BIC}_{\rm e\neq0} = 1.5$, implying that the circular and eccentric orbital solutions are almost equally favored. 
In other words, no evidence for non-zero eccentricity is found in our data set. 
A near-zero orbital eccentricity is also expected from the tidal circularization timescale for USP planets; using Equation (17) of \citet{2017AJ....154....4P} with 
the planetary tidal quality factor of $Q_p\approx 100$ (for a terrestrial planet) 
\citep[e.g.,][]{2021AJ....161...23M}, we obtain the tidal damping timescale of 
$\approx 5.5\times 10^4$ years for TOI-1634b, implying that a non-zero eccentricity
should have been damped in the past. 
Therefore,
we concluded that the TOI-1634b has an almost circular orbit, and adopt the fitting 
result for $e=0$ in the subsequent analysis. 
The RV data and the best-fit orbital solution to the data are plotted in panels (a) and (b) of Figure \ref{fig:toi1634_rv}.

The best-fit RV acceleration $\dot{\gamma}$ is $\approx 3$ times larger than the value
in the right hand side of Equation (\ref{eq:3}), 
but it is consistent with zero within $2-\sigma$. 
While this possibly large RV drift might be attributed to 
a moderate eccentricity of the binary orbit as discussed above, it could be an artifact caused by a small number of RV points around the beginning and/or end of our observing campaign spanning $\sim 5$ months. Given the frequency of the planet-multiplicity for USP planets \citep{2018NewAR..83...37W}, 
it is also possible that there exists an outer planet in the system
that gave systematic offsets at specific orbital phases for the inner USP planet. 
To discuss the significance of this RV trend, 
we next fitted the observed RV data in the absence of the RV trend $\dot{\gamma}$
assuming a circular orbit. Our MCMC analysis suggested $K=11.80\pm 0.91$ m s$^{-1}$, 
which is compatible with the result in the presence of $\dot{\gamma}$. 
Comparing the BIC's for the two fitting results, we found that
the result without the trend is equally likely ($\Delta\mathrm{BIC} = \mathrm{BIC}_{{\rm with}~\dot{\gamma}}-\mathrm{BIC}_{{\rm no}~\dot{\gamma}} = 0.6$). 
We thus list both fitting results (with and without $\dot{\gamma}$) in Table \ref{hyo3} 
to take into account the uncertainty of the systematic RV offset. 
We employ the result with $\dot{\gamma}$ and $e=0$, which is physically motivated from the
dynamics of the system, in deriving the planet mass $M_p$ 
as well as the mean density $\rho_p$ from $K$ (Table \ref{hyo2}).

After removing the best-fit orbital model ($e=0$, $\dot{\gamma}\neq 0$) for 
the observed RV data, we performed an extra periodogram analysis to search for 
additional planets in the system. 
The bottom panel of Figure \ref{fig:rv-period} illustrates the GLS periodogram 
(red solid line) for the residual 
RV data. No significant peak was found in the residual RVs, suggesting either that 
no additional massive planet is present in the system with the period shorter than our 
observation span or that the signal of such unidentified planets was removed/minimized 
by the orbital fit of TOI-1634b and long-term RV trend. 
At this point, our RV data imply no evidence for additional planets in the system.

\subsubsection{TOI-1685}

\begin{figure*}
\centering
\includegraphics[width=15cm]{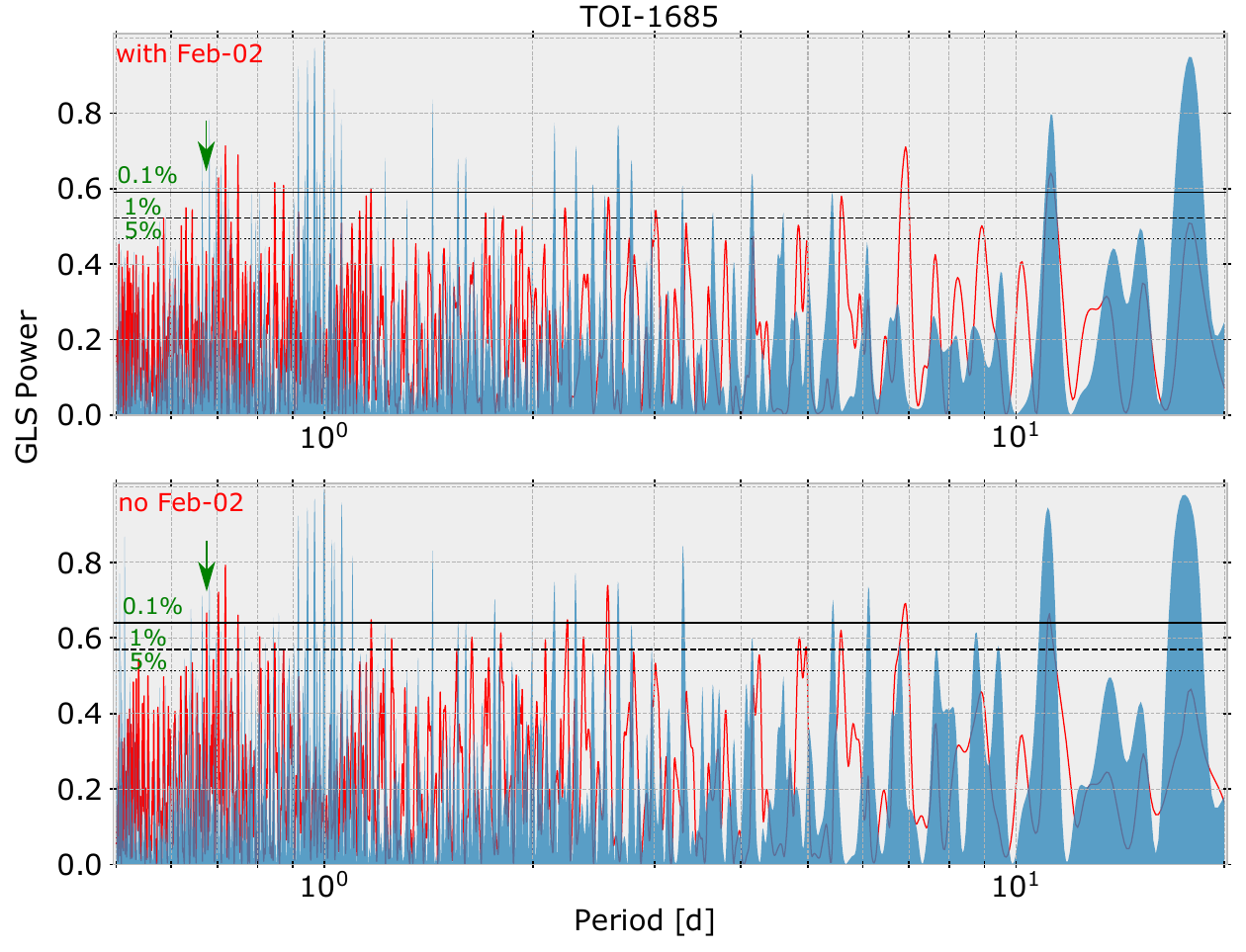}
\caption{
GLS periodograms (red solid lines) for TOI-1685's RV data with (upper panel) and without (lower panel) the Feb-02 data. As in Figure \ref{fig:rv-period}, blue shaded areas indicate the window function.
The black horizontal lines correspond to FAPs indicated in the plot. 
The period of TOI-1685.01 is denoted by the green arrow in both panels. 
}
\label{fig:rv-period2}
\end{figure*}

We ran a period analysis for the observed RV of TOI-1685 in a similar manner to TOI-1634.
The upper panel of Figure \ref{fig:rv-period2} plots the GLS periodogram for the raw RV data. 
There are several significant peaks exceeding the $\mathrm{FAP}=0.1\%$ line, but 
the one at the period of TOI-1685b ($P=0.669$ day) is not high enough to claim the detection of the
orbital signal. 
After a preliminary orbital fit to the observed RV data using the transit ephemeris, 
we found that the RV points taken on UT 2021 February 2 (hereafter, ``Feb-02")
are the primary outliers,
deteriorating the fitting result for the planet. 
Although this could be indicative of the presence of an additional planet in the system, 
we also suspected that this sudden RV shift is caused by an instrumental systematic. 
The IRD spectrograph is known to exhibit a relatively large temporal RV drift, which 
is well correlated with the temperature instability at the camera lens inside the chamber
\citep{2018SPIE10702E..11K, 2020PASJ...72...93H}. 
This instrumental RV drift is usually corrected by modeling the instantaneous
IP of the spectrograph derived from the simultaneously 
taken wavelength-reference spectrum (i.e., LFC). 
However, if the variation in IP is too fast compared to each integration time, 
it is theoretically expected that the LFC is unable to accurately trace the ``effective" instantaneous
IP of the spectrograph.

\begin{figure}
\centering
\includegraphics[width=8.5cm]{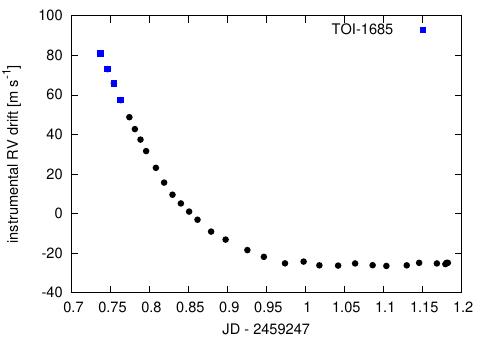}
\caption{
Temporal RV drift of IRD spectra on UT 2021 February 2, measured based on the emission lines of the LFC spectra. This apparent RV drift is caused by the temperature instability of the spectrograph. 
}
\label{fig:drift}
\end{figure}

To further investigate this possibility, we inspected the absolute instrumental drift 
of the spectrograph on February 2, and found that the IRD spectrograph 
indeed exhibits a large instrumental instability that night as shown in Figure \ref{fig:drift}. 
In particular, TOI-1685 was observed at the very beginning of the night (blue squares), 
when the instrumental RV variation was the most significant; the spectrograph
exhibits an RV drift of $\approx 8 $ m s$^{-1}$ for every integration\footnote{In most cases, 
the instrumental RV drift of IRD is less than $10-20$ m s$^{-1}$
over a few hours, but that night showed a particularly drastic RV variation during the first half night. }. 
In addition, the observing condition during the twilight usually changes dramatically, 
and thus the combination of the instrumental instability and variations in the twilight 
observing conditions may have affected the extraction and application of the effective
IPs from the LFC spectra.

The impact of IRD's instrumental RV drift, especially for the case of 
relatively long integrations, is under investigation, and therefore we decided to perform
the orbital fits with and without including the Feb-02 data. 
We first computed the periodogram for the data set excluding the Feb-02 data. 
The lower panel of Figure \ref{fig:rv-period2} plots the resulting GLS periodogram. 
While the same peaks ($\mathrm{FAP}<0.1\,\%$) identified for the original data set 
(upper panel) have similar GLS powers, the peak at the correct period of 
TOI-1685b ($P=0.669$ day) now appears with a low FAP ($<0.1\,\%$); 
whether instrumental or astrophysical, the absence of the significant peak 
at TOI-1685b's orbital period in the original periodogram is ascribed to 
the inclusion of the Feb-02 data. 
The two peaks around 0.70 day and 0.72 day in Figure \ref{fig:rv-period2}, which are higher than the $0.67-$day peak, are likely aliases associated with the peak at $2.59$ days. The window function has peaks at 1.0 day and 0.96 day (the highest and
second highest ones for $P<10$ days). 
When those window frequencies are coupled with the period at $2.59$ days, the periodogram would exhibit alias peaks around 0.72 and 0.70 day, respectively. 
The $2.59-$day periodicity will be discussed later.

\begin{figure}
\centering
\includegraphics[width=8.5cm]{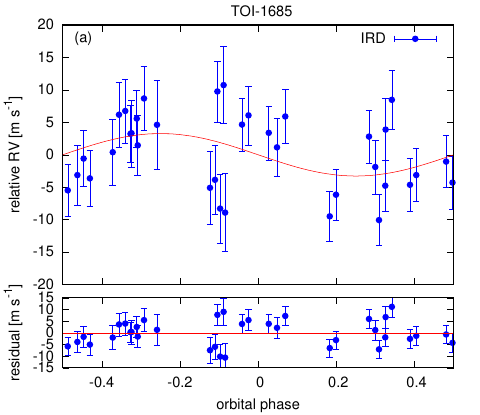}
\includegraphics[width=8.5cm]{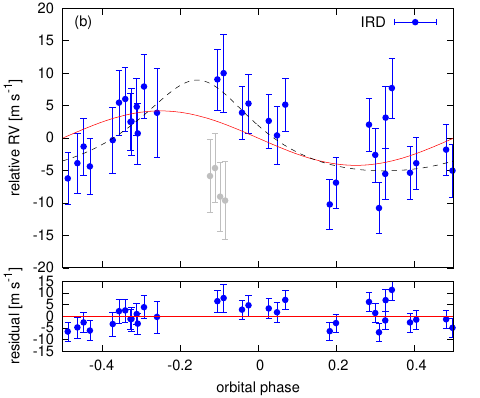}
\caption{
Results of the RV fits for TOI-1685 with a single-planet model
(a) with and (b) without the inclusion of the Feb-02 data. 
The blue points are observed RV data, and the red solid line indicate the 
best-fit circular model in each panel. In panel (b), we show the best-fit
eccentric orbit by the dashed line.  
In both panels, the RV residuals from the best-fit circular orbit are plotted
at the bottom. 
}
\label{fig:toi1685_rv}
\end{figure}

For the RV data with and without the Feb-02 data, 
we next fitted the observed RVs with a single-planet model. 
Assuming either a circular or eccentric orbit, we performed the MCMC analysis
as in the case of TOI-1634 for each data set. 
When the Feb-02 data were included, we obtained $K=3.3\pm1.1$ m s$^{-1}$
and $K=4.1_{-1.4}^{+1.5}$ m s$^{-1}$ for the circular and eccentric orbits, respectively.
The two fitting results yielded 
$\Delta\mathrm{BIC}= \mathrm{BIC}_{\rm e=0}-\mathrm{BIC}_{\rm e\neq0} = -2.3$, 
implying that the circular orbit is slightly favored for this data set. 
We obtained larger $K$ values in the absence of the Feb-02 data:
$K=4.2\pm1.1$ m s$^{-1}$ and $K=7.0_{-1.6}^{+1.5}$ m s$^{-1}$ for $e=0$ and $e\neq 0$, respectively. 
In this case, the two fits resulted in 
$\Delta\mathrm{BIC}= \mathrm{BIC}_{\rm e=0}-\mathrm{BIC}_{\rm e\neq0} = +4.0$;
unlike the case with the Feb-02 data, an eccentric orbit is a slightly favorable solution. 
Note that as in the case of TOI-1634, the tidal circularization timescale for 
TOI-1685.01 is estimated as $\approx 1.0\times 10^4$ years for $Q_p\approx 100$ 
(Earth-like rocky planet), indicating that $e$ should be vanishingly low in the 
absence of an additional planet in the system. 
Those fitting results are shown in Table \ref{hyo3} and 
the phase-folded RVs are plotted in Figure \ref{fig:toi1685_rv}. 
For the final planet mass $M_p$ (Table \ref{hyo2}), 
we adopt the $K$ value for the case of $e=0$ without the Feb-02 data.

\begin{figure*}
\centering
\includegraphics[width=15cm]{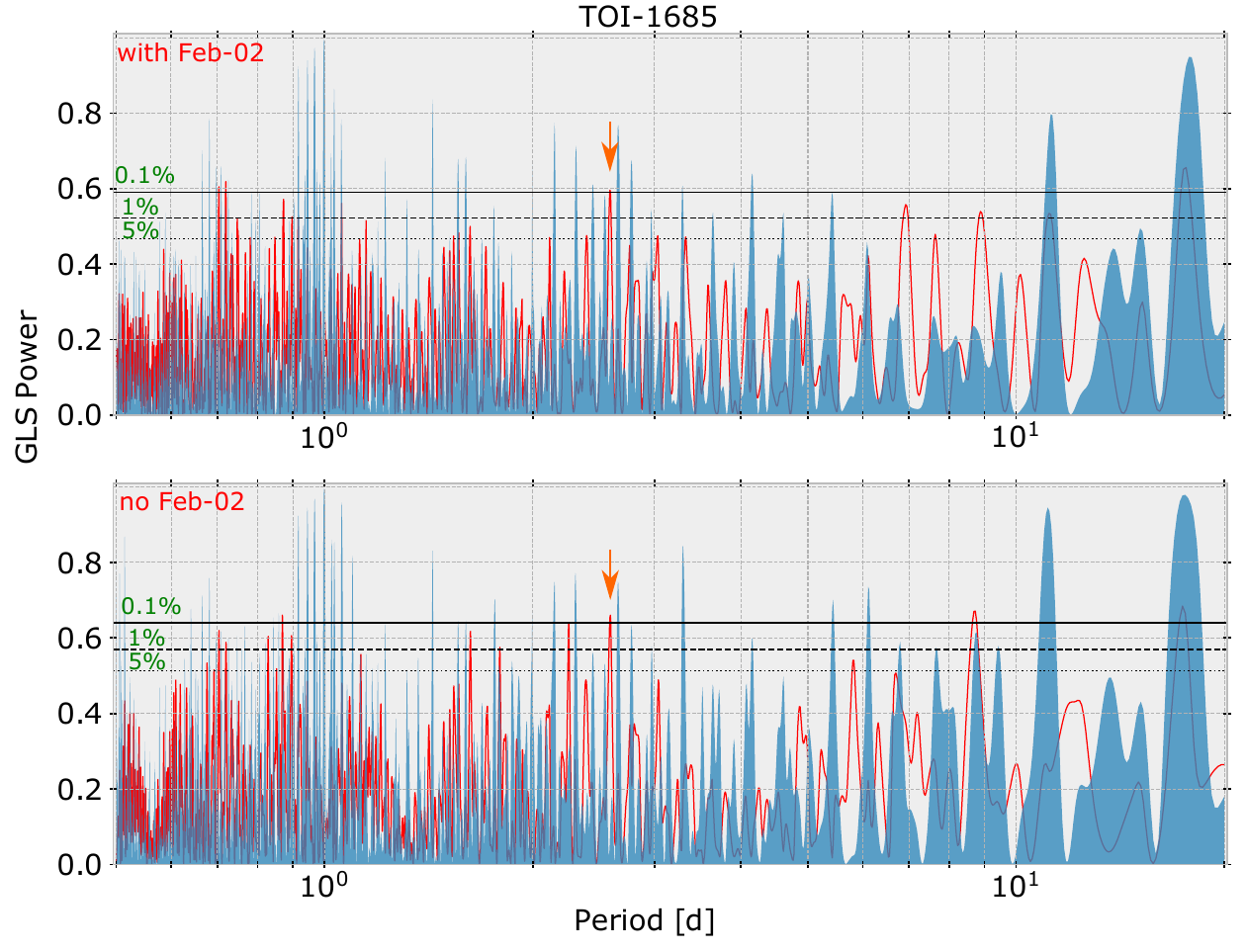}
\caption{
GLS periodograms computed for TOI-1685's residual RV data after subtracting 
the best-fit Keplerian motion by the USP planet (TOI-1685.01). 
The results with and without the Feb-02 data are shown in the upper and lower
panels, respectively. 
The $2.6-$day periodicity discussed in the text is shown by the orange arrow
in each panel. 
}
\label{fig:rv-period3}
\end{figure*}

In order to search for an additional signal in the observed RV data, we computed 
the periodogram for TOI-1685's RVs after removing the best-fit single-planet model
for each data set. 
Considering the short tidal circularization timescale for the USP planet,
we removed the circular-orbit solutions derived 
above. Figure \ref{fig:rv-period3} plots the GLS periodograms for the whole RV data
and the data subset without the Feb-02 data. 
For both panels, there are a few significant peaks ($\mathrm{FAP}<0.1\%$) that do not
fall in the window function. The peak at $2.6$ days is common to both periodograms, 
which was also seen in the original RVs without the Feb-02 data 
(lower panel of Figure \ref{fig:rv-period2}). 
The high peaks at $P<1.0$ day in both panels are likely alias peaks associated 
with the $2.6-$day peak and window functions.

\begin{table}[t]
\centering
\caption{Results of the Orbital Fits for TOI-1685 with a Two-planet Model}\label{hyo4}
\begin{tabular}{lcc}
\hline\hline
Parameter & with Feb-02 Data & no Feb-02 Data \\\hline
$K_1$ (m s$^{-1}$) & $4.8_{-1.2}^{+1.1}$ & $4.9_{-1.5}^{+1.3}$  \\
$K_2$ (m s$^{-1}$) & $6.2\pm 1.0$ & $5.6_{-1.0}^{+1.0}$ \\
$P_2$ (days) & $2.5909_{-0.0048}^{+0.0045}$ & $2.5891_{-0.0069}^{+0.0054}$  \\
$T_{c,2}$ (BJD$_{\rm TDB}$) & $2458827.24_{-0.64}^{+0.67}$ & $2458827.46_{-0.75}^{+0.90}$ \\
\hline
\end{tabular}
\end{table}

\begin{figure}
\centering
\includegraphics[width=8.5cm]{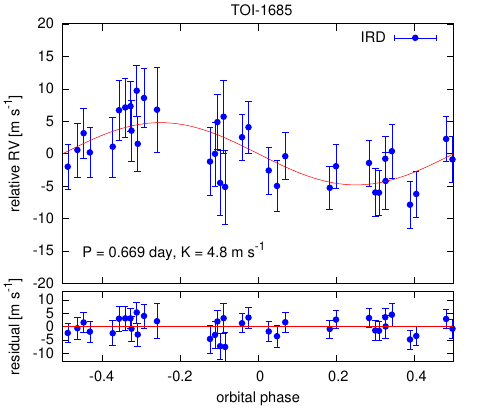}
\includegraphics[width=8.5cm]{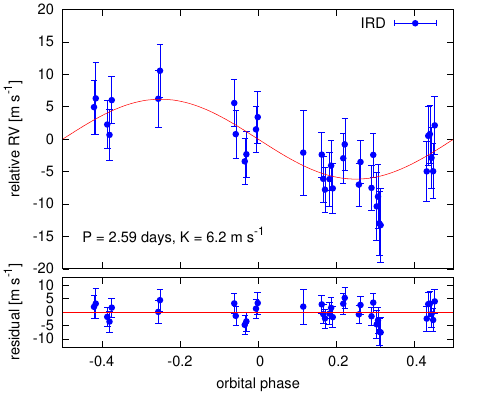}
\caption{
The result of the RV fit for TOI-1685 with a two-planet model. 
Phase-folded RV curves for the USP planet (upper panel) and the outer one 
(lower panel) are respectively shown after subtracting the best-fit Keplerian orbit 
for the other planet. 
}
\label{fig:toi1685_twopla}
\end{figure}

Given the limited phase coverage and unknown instrumental systematics, at this
point we are not able to claim that the $2.6-$day periodicity in the RV data represents
an additional planet in the system; more RV measurements are essentially required
to gain a robust conclusion on the presence of an additional body in the system. 
Nonetheless, we were tempted to fit the observed RV data with a two-planet model. 
In doing so, we ran the MCMC code and fitted the RV data (with and without 
the Feb-02 data) assuming two circular Keplerian orbits. 
We fixed the period of the USP planet at the one from the transit ephemeris and 
allowed the period of the outer planet $P_2$ and time of the inferior conjunction 
$T_{c,2}$ to float with uniform priors. 
The results of these fits are listed in Table \ref{hyo4}. In the table, $K_1$ and $K_2$
represent the RV semi-amplitudes for the inner (USP) and outer planets, respectively. 
The phase-folded RV curves (with the inclusion of Feb-02 data) after removing the Keplerian orbit for the other planet are shown in Figure \ref{fig:toi1685_twopla}. 
The RV semi-amplitudes for the USP planet are consistent within $\approx 1\,\sigma$ 
with the values derived for the one-planet model (Table \ref{hyo3}) in both cases,
whereas the RV scatters around the best-fit models significantly improved
with $\Delta\mathrm{BIC}=\mathrm{BIC}_\mathrm{one-planet}-\mathrm{BIC}_\mathrm{two-planet}$ being greater than 10 for both fits.

We note that the $2.6-$day signal is unlikely to be explained by stellar rotation.
If the rotation period of the star is $P_{\rm rot}=2.6$ days, 
the equatorial rotation velocity
must be $\approx 8.8$ km s$^{-1}$, which also gives the projected rotation velocity 
$v\sin i_s$ for the case of spin-orbit alignment in the system. 
Both TRES optical spectra and IRD near IR spectra, however, imply that
the star is slowly rotating with $v\sin i_s<5$ km s$^{-1}$. 
The slow rotation of TOI-1685 is also supported by the low-frequency light 
curve modulation as discussed in Section \ref{sec:photometry}. 
Therefore, we conclude that the $2.6-$day periodicity does not indicate the
rotational signal in the RV data, but represents any one of
(1) an additional planet, (2) an instrumental/analysis artifact, (3) an artifact 
caused by the mixture of (1) and (2) as well as the window function of our IRD
observations. 
Again, further observations are required to test on those possibilities.

If the $2.6-$day signal indeed represents the period of the outer planet, 
$K_2\approx 6$ m s$^{-1}$ corresponds to the planetary mass of 
$M\sin i_o=7-8\,M_\oplus$. 
Although the two planets in the system have relatively small masses, 
the small orbital separation between the two planets prompted us to check 
for the orbital stability of the two planets. Since the outer one is not transiting
and its orbital inclination (thus the true mass) is not known, currently 
there is little point in running detailed numerical simulations for the system. 
Instead, we simply compared the minimum separation between the two
in terms of the mutual Hill sphere $R_H$, 
following \citet{2015ApJ...807...44P}. 
Inputting the semi-major axes of the two planets ($a_1=0.0116$ au 
and $a_2=0.0285$ au, for the inner and outer planets, respectively)
on the assumption that the planets are coplanar, we found $R_H\approx 0.00058$ au. 
Thus, the minimum separation between the two planets ($a_2-a_1=0.0169$ au)
is about 29 times larger than the mutual Hill radius. 
\citet{2015ApJ...807...44P} showed that if the minimum separation is larger than 
$\approx 12\,R_H$, the system should be stable on a billion-year timescale. 
Also considering that the periods of the two planets are not near a first-order mean motion
resonance, the addition of a super-Earth-mass planet at $P=2.6$ days
does not critically deteriorate the stability of the system.


\section{Discussion} \label{sec:discussion}

\subsection{Planet Compositions}
\begin{figure*}
\centering
\includegraphics[width=15cm]{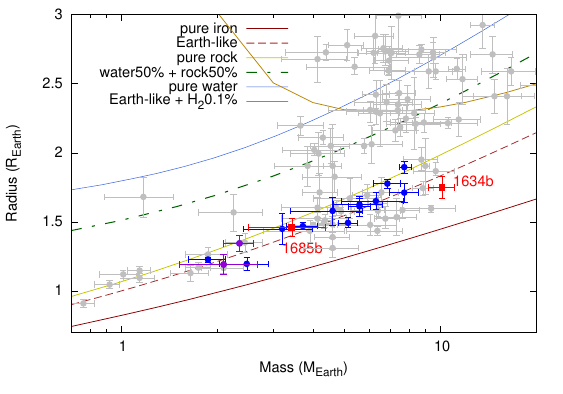}
\caption{
MR diagram for known transiting planets ($R_p<3.0\,R_\oplus$) 
as well as our new planets (blue squares). 
The catalog was downloaded from the TEPcat database \citep{2011MNRAS.417.2166S}
and theoretical models are drawn based on \citet{2016ApJ...819..127Z, 2019PNAS..116.9723Z}. 
USP planets around M dwarfs (except our new planets) and FGK stars are shown in purple and blue, respectively. 
}
\label{fig:MR}
\end{figure*}

Based on the results of light curve analyses and RV fits, we estimated the physical
parameters of the planets, including the mass $M_p$, radius $R_p$, semi-major axis $a$, and equilibrium temperature $T_{\rm eq}$ assuming zero albedo ($A_B=0$) as well as Earth-like albedo ($A_B=0.3$), which are listed in Table \ref{hyo2}. 
In computing $T_{\rm eq}$, we assumed a constant temperature across the entire planet. 
To plot the two planets in the mass-radius (MR) diagram for exoplanets, we downloaded the catalog of transiting planets from the TEPcat database \citep{2011MNRAS.417.2166S} and used the mass and radius of well-characterized planets, with the precisions on both measurements better than $30\,\%$. 
Figure \ref{fig:MR} shows the MR diagram, focusing on relatively small-sized planets with $R_p<3.0\,R_\oplus$. 
The blue and purple points in the figure indicate the USP planets in the literature, while the gray ones are other longer-period planets. 
In the same figure, MR curves for different planet compositions are drawn based on 
the theoretical MR relations by \citet{2016ApJ...819..127Z, 2019PNAS..116.9723Z}. 
For models including water and/or hydrogen atmosphere, a surface temperature of 1000 K
is assumed in the plot based on the equilibrium temperature of the planets in Table \ref{hyo2}. 
Models including water-rich cores with hydrogen envelopes are not shown in the figure, as the radii of such planets usually exceed $3.0\,R_\oplus$ even with the smallest 
addition of hydrogen envelope (i.e., $0.1\,\%$ of H$_2$).

The derived mean densities for TOI-1634b and TOI-1685b are 
$10.4_{-1.6}^{+1.9}$ g cm$^{-1}$ and $6.1_{-1.7}^{+1.9}$ g cm$^{-1}$, respectively,
which are higher than that of Earth. 
All the USP planets plotted in Figure \ref{fig:MR} including our new planets TOI-1634b and TOI-1685b have interior compositions consistent with Earth's composition (i.e., $32.5\,\%$ Fe + $67.5\,\%$ MgSiO$_3$) or pure rock (which is only allowed for TOI-1685b), and the diagram implies that it is very unlikely that the two planets possess light element (H-He) rich atmospheres. 
Among the USP planets plotted in the diagram, TOI-1634b is one of the largest and
most massive planets having Earth-like compositions. 
The radius of TOI-1634b falls near the radius gap of super-Earths
\citep{2017AJ....154..109F}, which makes the planet a benchmark for a population of
large USP planets around low-mass stars; residing near the radius gap,
TOI-1634b is useful in the context of discussing to what extent the rocky cores of close-in planets can grow and how such large planets were delivered to the present locations and lost their volatile-rich envelopes.
TOI-1685b is more like a typical USP planet with $R_p\lesssim 1.5\,R_\oplus$,
whose composition is consistent with Earth.

\subsection{Atmospheric Escape from the USP Planets}\label{sec:photo-eva}
Our finding that both TOI-1634b and TOI-1685b are almost ``bare" planets
having little, if any, volatile-rich atmosphere is corroborated in the context of 
the photo-evaporation theory, independently of the observed mean densities.
Atmospheric escapes are generally driven by several physical processes
\citep[e.g.,][]{Tian2015}.
USP planets having massive atmospheres are in danger of tidal disruption.
If TOI-1685\,b initially had a primordial atmosphere of $\gtrsim 10-20$\,\% of its total mass at the current location, the atmosphere should have been blown off instantaneously by the Roche lobe overflow because of its small core mass and a high equilibrium temperature, whereas
the more massive TOI-1634b has never experienced the Roche lobe overflow if it initially had such a massive atmosphere.
The observed mass-radius relationship, however, rules out the presence of such a massive atmosphere on the two USP planets. 

The primordial atmosphere on a USP planet is exposed to an intense stellar irradiation and high-energy, charged particles from a stellar wind and coronal mass ejection. 
In particular, the hydrodynamic escape driven by high-energy (X-ray and extreme UV: XUV) photons from the host star \citep[e.g.,][]{Sekiya1980,Watson1981} plays a crucial role for highly-irradiated close-in planets \citep[]{Owen2019}.
We simulated the long-term evolution of TOI-1634\,b and 1685\,b that initially have the atmospheric mass fraction of $\lesssim$ a few \% on an Earth-like core under a strong stellar XUV irradiation.
We used the physical properties of the two USP systems given in Tables \ref{hyo1} and \ref{hyo2}.
We adopted the XUV flux model for M-dwarfs given in \citet[][]{2012MNRAS.422.2024J}, where the bolometric luminosities of TOI-1634 and 1685 were assumed to be their current values.
The hydrodynamic mass loss from a planet with a H$_2$-He atmosphere is calculated by 
\begin{equation}
  \dot{M_\mathrm{p}} = - \eta \frac{{R^3_\mathrm{p}} L_\mathrm{XUV}(t)}{4  G  M_\mathrm{p}  a^2  K_\mathrm{tide}},
 \end{equation}
where $\eta$ is the heating efficiency by a stellar XUV irradiation, $L_\mathrm{XUV}$ is the stellar XUV luminosity, $R_\mathrm{p}$ is the planetary radius, $a$ is the semi-major axis of the planet, $G$ is the gravitational constant, and $K_\mathrm{tide}$ is the potential energy reduction factor due to the stellar tidal effect \citep{2007A&A...472..329E}. We adopted $\eta = 0.1$ for low-mass planets as suggested in \citet{2012MNRAS.425.2931O}. The planetary radius, which is defined as the location at which a H$_2$-He atmosphere becomes optically thick to stellar XUV photons, can be determined by the thermal evolution of the planet (see also \citet[][]{2020ApJ...889...77H} for detailed numerical prescriptions).

The two USP planets are expected to completely lose their primordial (i.e., H- and He-rich) atmospheres by photoevaporation within $\lesssim 1$\,Gyr,
which are also consistent with the mass loss timescales given in \citet[]{Owen2017} (see equation (20)). 
Although the precise ages of TOI-1634 and TOI-1685 are not well-determined, 
the two USP systems exhibit no sign of particular youth (e.g., rapid rotation and high surface activity). 
Thus, once the USP planets accreted a primordial atmosphere of $\lesssim$ a few wt\,\% from the protoplanetary disk,
all the atmospheres are likely to be lost by photoevaporation processes.
Hence, 
both planets should be bare planets, similarly to the other known USP planets.

\subsection{Further Follow-up Studies}

As discussed above, TOI-1685's RV data exhibit an extra scatter when fitted with
a single-planet model, which could be attributed to the presence of an additional 
planet or activity/instrument induced systematic effects. 
More RV observations are beneficial not only to confirm or rule out the presence of
an outer planet, but also to obtain an accurate mass for the USP planet;
in the presence of unknown additional bodies in the system, the mass measurement
of a known transiting planet is more or less affected by the systematic variations due to additional bodies. Therefore, the planet masses given in Table \ref{hyo2} and 
Figure \ref{fig:MR} are tentative ones, whose systematic errors might be underestimated. 
We note that in many theoretical scenarios of the USP planet formation, outer planets 
play a key role in bringing the USP planets to the current locations
\citep[e.g.,][]{2010ApJ...724L..53S, 2019MNRAS.488.3568P, 2020ApJ...905...71M}.
Additional RV monitoring would be able to uncover the architecture of the two systems
up to a larger orbital distance. Specifically, once an outer planet is confirmed beyond
the USP planet, properties such as the period ratio, orbital eccentricity, and 
mutual inclinations would be valuable clues to test the formation scenarios of
USP planets. 

The brightness of TOI-1634 and TOI-1685 also makes them excellent targets for future 
follow-up studies, including atmospheric characterizations. 
As discussed in Section \ref{sec:photo-eva}, the primordial H-He atmosphere 
of the UPS planets would have been lost due to strong irradiation of high-energy 
photons from the host stars. However, the planets may still hold a geometrically 
thin layer of
atmosphere comprised of heavy elements, formed e.g., by a constant outgassing from the 
planet interior \citep[e.g.][]{2018A&A...614A..18D} or degassing from accreted material such as meteorites \citep[e.g.][]{2008ApJ...685.1237E}. In order to bring TOI-1634b and TOI-1685b 
in the context of atmospheric characterizations, 
either by emission spectroscopy or transmission spectroscopy, 
we calculated the emission spectroscopy metric (ESM) as well as the transmission 
spectroscopy metric (TSM) for TOI-1634b and TOI-1685b, 
introduced by \citet{2018PASP..130k4401K}. 
In short, those metrics allow us to compare the relative observational signals for
atmospheric characterizations based on the intrinsic strength of the spectroscopic 
features and the target apparent magnitude. 
ESM measures the expected signals mainly for secondary eclipse observations, 
while TSM is an index for transmission spectroscopy.

We downloaded the stellar and planetary parameters for known planetary systems from 
the NASA exoplanet 
archive\footnote{https://exoplanetarchive.ipac.caltech.edu/index.html}, 
and extracted transiting planets with measured masses. Planet masses are required
for TSM since TSM depends on the scale height of the planet atmosphere. 
We computed Equations (1) and (4) of \citet{2018PASP..130k4401K}, 
in which we input the equilibrium temperatures
of planets assuming $A_B=0$; since we are only interested in the ``relative"
observational signals, non-zero albedos have no impact on the overall rank order.  
The $J-$band magnitudes (required for TSM) were not available in the 
downloadable table of the
Exoplanet archive catalog, and thus we adopted the values from 2MASS \citep{2006AJ....131.1163S} by matching the target names or coordinates
via the SIMBAD database \citep{2000A&AS..143....9W}. 
Following the prescription in \citet{2018PASP..130k4401K}, we changed the
``scale factor" for the TSM with the planet radius, which depends on the mean
molecular weight $\mu$ of the exoplanet atmosphere. 
\citet{2018PASP..130k4401K} changed the scale factor only by the planet radius,
but we also took into
account the mean density of each planet; for the planets with $R_p<1.5\,R_\oplus$, 
we fixed the scale factor at 0.19 (as they are almost always rocky planets), and for those with $R_p\geq 1.5\,R_\oplus$, 
it was set to 0.19 and 1.26 \citep{2018PASP..130k4401K} when the mean density 
is higher and lower than that of the Earth, respectively. 
In calculating the two metrics, we focused on transiting planets with 
$R_p<2.0\,R_\oplus$ so as to compare the relative atmospheric signals for small,
mostly rocky planets, whose compositions are similar to that of Earth. 

\begin{figure}
\centering
\includegraphics[width=8.5cm]{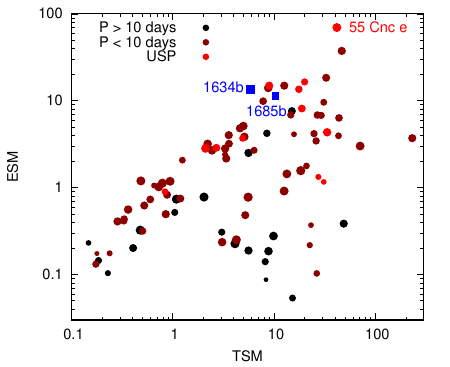}
\caption{ESM and TSM \citep{2018PASP..130k4401K} are plotted for all transiting planets ($R_p<2.0\,R_\oplus$) with measured masses. 
Those metrics indicate the relative signals of atmospheric characterizations 
by emission (ESM) and transmission spectroscopies (TSM). 
The size of the symbols corresponds to the planet radius. 
}
\label{fig:metrics}
\end{figure}

Figure \ref{fig:metrics} plots the two metrics for all known transiting planets ($R_p<2.0\,R_\oplus$) as well as our newly confirmed planets. 
According to ESM, both TOI-1634b and TOI-1685b are ranked in the top ten best targets
for emission spectroscopy, while they are ranked moderately high for transmission spectroscopy; 
planets around mid-to-late M dwarfs (e.g., TRAPPIST-1) smaller than TOI-1634 and
TOI-1685 are better suited for atmospheric characterizations by transmission spectroscopy thanks to the enhanced transit depths. TOI-1634b and TOI-1685b are more favorable targets for observations of secondary eclipses. 
It should be noted that the top 3 ranked planets according to ESM 
(55 Cnc e, HD 219134b, and HD 219134c)
are probably not suitable for emission spectroscopy using large-aperture telescopes 
\citep[e.g., JWST;][]{2014PASP..126.1134B} since their host stars are too bright ($K_s<5$ mag) for efficient observation with a large telescope.

\section{Summary} \label{sec:summary}

With a goal of confirming and characterizing the USP planet candidates around TOI-1634 
and TOI-1685, we conducted intensive follow-up observations for the two targets 
including ground-based transit photometry, high-resolution imaging, reconnaissance 
spectroscopy, and high-precision RV measurements. 
The light curves from the ground-based photometry indicated the transit depths consistent with those by the TESS photometry. 
The ground-based photometry also helped us to refine the 
orbital periods by more than an order of magnitude compared with the
ephemeris obtained from TESS photometry alone.  
Spectroscopic follow-up observations revealed that the two stars are both
$\approx\,\mathrm{M3}$ dwarfs on the main sequence, having very similar effective temperatures, masses, and radii.

TOI-1634 has a bound, low-mass companion star separated by $2\farcs5$, 
which is estimated to have $\approx 0.1\,M_\odot$, whose flux is contaminated in 
the light curves by 
both TESS and ground-based photometric observations, but we confirmed that 
the USP planet is indeed orbiting the primary star through the RV measurements 
using Subaru/IRD;
the periodogram for the RV data exhibits the highest peak at the period of TOI-1634b
($0.989$ day), and its observed variations indicated the USP planet has a mass of 
$M_p=10.14 \pm 0.95\,M_\oplus$ when a circular orbit is assumed.

On the other hand, TOI-1685's RV data show a more puzzling behavior, but a significant 
peak ($\mathrm{FAP}<0.1\,\%$) is detected at the right transit period ($0.669$ day) 
in the periodogram when the RV points taken during a significant instrumental
instability (Feb-02) were removed from the analysis. As a result of fitting 
the observed RVs without the Feb-02 data, we obtained the USP planet mass of 
$3.43 \pm 0.93\,M_\oplus$ for the case of a circular orbit. 
The residual RVs around the best-fit circular model show an excess scatter, 
suggesting the presence of a moderate eccentricity, an unknown systematic effect by instrumental or stellar-activity induced noise, and/or an additional planet in the system; the secondary-planet scenario is the most preferred scenario according to the
BIC values for different scenarios. 
The additional periodogram analysis on the residual RVs indeed suggests a possible
periodicity at $\approx 2.6$ days, but we were unable to claim that it is 
an additional planet signal due to the lack of phase coverage and unknown nature 
of stellar activity.  
Further observations are needed to confirm the presence of the additional planet(s).

When the planet masses for the circular, one-planet model are adopted 
(Table \ref{hyo2}), both TOI-1634b and TOI-1685b are plotted near the theoretical curve
for the Earth-like composition in the MR diagram. Therefore, the two new USP planets
should have similar properties to those of all the other USP planets with
$R_p<3\,R_\oplus$, including the internal structure and atmosphere. 
TOI-1634b is one of the largest and most massive USP planets having an Earth-like
composition, and therefore, would become a benchmark target to study the formation 
and evolution history of massive USP planets. 
Both planets are listed among the best suitable targets for future atmospheric
studies of small rocky planets by emission spectroscopy thanks to the brightness of 
the host stars, which encourages future characterizations using large aperture
telescopes including JWST. 
Although small USP planets ($<2\,R_\oplus$) are likely to have lost the primordial 
atmospheres dominated by H$_2$ and He, 
one may be able to probe and constrain the secondary atmosphere formed via
the outgassing from the planet interior.

\acknowledgments
This work is partly supported by JSPS KAKENHI Grant Numbers JP20K14518, JP19K14783, JP18H01265, JP18H05439, JP17H04574, JP18H05442, JP15H02063, JP21H00035, and JP22000005, JST PRESTO Grant Number JPMJPR1775, Grant-in-Aid for JSPS Fellows, Grant Number JP20J21872, 
and a University Research Support Grant from the National Astronomical Observatory of Japan (NAOJ).
JNW thanks the Heising-Simons Foundation for support.
The data analysis was carried out, in part, on the Multi-wavelength Data Analysis System operated by the Astronomy Data Center (ADC), National Astronomical Observatory of Japan.

Based on observations obtained at the Observatoire du Mont-M\'{e}gantic, financed by Universit\'{e} de Montr\'{e}al, Universit\'{e} Laval, the National Sciences and Engineering Council of Canada (NSERC), the Fonds qu\'{e}b\'{e}cois de la recherche sur la Nature et les technologies (FQRNT), and the Canada Economic Development program and the Quebec Minist\`{e}re de l'\'{E}conomie et de l'Innovation. 

This paper includes data collected by the \textit{TESS} mission. Funding for the \textit{TESS} mission is provided by the NASA Explorer Program. We acknowledge the use of \textit{TESS} Alert data, which is currently in a beta test phase, from pipelines at the \textit{TESS} Science Office and at the \textit{TESS} Science Processing Operations Center. Resources supporting this work were provided by the NASA High-End Computing (HEC) Program through the NASA Advanced Supercomputing (NAS) Division at Ames Research Center for the production of the SPOC data products. This research has made use of the Exoplanet Follow-up Observation Program website, which is operated by the California Institute of Technology, under contract with the National Aeronautics and Space Administration under the Exoplanet Exploration Program. 

This work has made use of data from the European Space Agency (ESA) mission {\it Gaia} (\url{https://www.cosmos.esa.int/gaia}), processed by the {\it Gaia} Data Processing and Analysis Consortium (DPAC, \url{https://www.cosmos.esa.int/web/gaia/dpac/consortium}). Funding for the DPAC has been provided by national institutions, in particular the institutions participating in the {\it Gaia} Multilateral Agreement.

This work was enabled by observations made from the Subaru, Gemini North, and Keck telescopes, located within the Maunakea Science Reserve and adjacent to the summit of Maunakea. We are grateful for the privilege of observing the Universe from a place that is unique in both its astronomical quality and its cultural significance.

Some of the Observations in the paper made use of the High-Resolution Imaging instrument 'Alopeke. 'Alopeke was funded by the NASA Exoplanet Exploration Program and built at the NASA Ames Research Center by Steve B. Howell, Nic Scott, Elliott P. Horch, and Emmett Quigley. Data were reduced using a software pipeline originally written by Elliott Horch and Mark Everett. 'Alopeke was mounted on the Gemini North telescope of the international Gemini Observatory, a program of NSF’s OIR Lab, which is managed by the Association of Universities for Research in Astronomy (AURA) under a cooperative agreement with the National Science Foundation on behalf of the Gemini partnership: the National Science Foundation (United States), National Research Council (Canada), Agencia Nacional de Investigación y Desarrollo (Chile), Ministerio de Ciencia, Tecnología e Innovación (Argentina), Ministério da Ciência, Tecnologia, Inovações e Comunicações (Brazil), and Korea Astronomy and Space Science Institute (Republic of Korea). Data collected under program GN-2020B-LP-105.

This paper is based on observations made with the MuSCAT2 instrument, developed by Astrobiology Center, at Telescopio Carlos Sánchez operated on the island of Tenerife by the IAC in the Spanish Observatorio del Teide.
This paper is based on observations made with the MuSCAT3 instrument, developed by the Astrobiology Center and under financial supports by JSPS KAKENHI (JP18H05439) and JST PRESTO (JPMJPR1775), at Faulkes Telescope North on Maui, HI, operated by the Las Cumbres Observatory.

This work makes use of observations from the LCOGT network. Part of the LCOGT telescope time was granted by NOIRLab through the Mid-Scale Innovations Program (MSIP). MSIP is funded by NSF.

%

\vspace{5mm}
\facilities{Subaru/IRD, Gemini-North/'Alopeke, Keck-II/NIRC2, LCOGT, Okayama-1.88m/MuSCAT, TCS/MuSCAT2, FTN/MuSCAT3}


\software{
{\tt AstroImageJ} \citep{Collins:2017}, 
{\tt IRAF} \citep{1993ASPC...52..173T}, 
{\tt SpecMatch-Emp} \citep{2017ApJ...836...77Y},
{\tt BANYAN $\Sigma$} \citep{2018ApJ...856...23G}, 
{\tt TAPIR} \citep{Jensen:2013}
}







\bibliographystyle{aasjournal}



\end{document}